\makeatletter
  \input{jbw-switch.sty}
\makeatother

\DeclareSwitch{long}
\DeclareSwitch{final}
\DeclareSwitch{color}
\DeclareSwitch{print}
\DeclareSwitch{jbw}
\IfFileExists{./jbw.tex}
  {\jbwtrue}
  {\jbwfalse}

\longfalse
\iflong
  \def\longonly#1{#1}%
\else
  \def\longonly#1{}%
\fi

\finalfalse

\colortrue

\printtrue

\documentclass[submission,creativecommons,noderivs]{eptcs}

\usepackage{jbw-switch}

\DeclareSwitch{color}

\ifcolor
  \usepackage[usenames]{color}
\else
  \usepackage[monochrome,usenames]{color}
\fi

\ifcolor
  \newcommand{\ColorA}{green}
  \newcommand{\ColorB}{red}

\else
  \newcommand{\ColorA}{black}
  \newcommand{\ColorB}{black}

\fi

\usepackage{pstricks}

\ifprint
  \definecolor{red}{cmyk}{0.00,0.80,0.94,0.00}
  \definecolor{green}{cmyk}{0.67,0.00,0.97,0.00}
  \definecolor{blue}{cmyk}{0.87,0.62,0.00,0.00}
  \definecolor{magenta}{cmyk}{0.29,0.55,0.00,0.00}
  \definecolor{gray}{cmyk}{0.42,0.35,0.31,0.00}
\else
  \definecolor{red}{rgb}{1.00,0.00,0.00}
  \definecolor{green}{rgb}{0.00,0.70,0.00}
  \definecolor{blue}{rgb}{0.00,0.00,1.00}
  \definecolor{magenta}{rgb}{1.00,0.00,1.00}
  \definecolor{gray}{rgb}{0.50,0.50,0.50}
\fi

\iffalse
  \usepackage[breaklinks=true,%
              dvips,%
              draft=false,%
              pdfborder={0 0 0},%
              pdfborderstyle={},%
              ]%
    {hyperref}
\else
\fi

  \usepackage[nowasy]{jbw-utf8}

\usepackage{proof}
\usepackage[nomargin,inline]{fixme}

\usepackage{jbw-iso-date}
\usepackage{jbw-math}
\usepackage{jbw-arrow}
\usepackage{jbw-array-space}
\usepackage{jbw-thm}
\usepackage{jbw-fix-hyperref-autoref}

   \usepackage{txfonts} 
\usepackage{pst-tree}
\psset{levelsep=0.7cm,treefit=tight,nodesep=2pt}

\DeclareRobustCommand{\mathtext}[1]
  {\ifmmode
     \mathpalette\mathtextaux{#1}%
   \else
     \mbox{#1}%
   \fi}
\newcommand{\mathtextaux}[2]
  {
   \mbox{\ifx#1\scriptstyle
           \scriptsize
         \else
           \ifx#1\scriptscriptstyle
             \scriptsize
           \fi
         \fi
         #2}}

\newcommand{\mtnorm}[1]{\mathtext{\textnormal{#1}}}
\newcommand{\mtnormsf}[1]{\mtnorm{\textsf{#1}}}

\newcommand{\upalpha}{{\mtnorm{\textalpha}}}

\newcommand{\upsflambda}{{\mtnormsf{\textlambda}}}
\newcommand{\ColorDef}{black}
\newcommand{\DefiningUse}[1]{\emph{\textcolor{\ColorDef}{#1}}}

\newcommand{\GrammarSym}{{\Coloneqq}}
\newcommand{\AppSym}{{\ensuremath\mathsf{@}}}

\newcommand{\FunTypeSym}{{\rightarrow}}

\newcommand{\IntsectSymAux}[2]
  {
   #1%
   \setbox0=\hbox{$\mathsurround=0pt\relax#1\cap$}%
   \mathbin
     {
      \raise0.4\ht0%
         \hbox to\wd0%
           {\hfill
            $\mathsurround=0pt\relax.$%
            \hfill}%
      \kern -\wd0\relax
      {\cap}}}

\newcommand{\ConstraintSymAux}[2]
  {
   #1%
   \setbox0=\hbox{$#1\lessdot$}%
   \mathbin
     {\kern 0.12\wd0\relax
      \vrule width 0.74\wd0 depth 0.5ex height -0.4ex\relax
      \kern 0.14\wd0\relax
      \kern -\wd0\relax
      {\lessdot}}}

\newcommand{\CircDotSymAux}[2]
  {
   #1%
   \setbox0=\hbox{$\mathsurround=0pt\relax#1\circ$}%
   \mathbin
     {
      \raise0.475\ht0%
         \hbox to\wd0%
           {\hfill
            $\mathsurround=0pt\relax.$%
            \hfill}%
      \kern -\wd0\relax
      {\circ}}}

\newcommand{\StuffX}{\metavar{X}}
\newcommand{\StuffY}{\metavar{Y}}
\newcommand{\StuffZ}{\metavar{Z}}
\newcommand{\Stuff}{\StuffX}

\newcommand{\FstSym}{\mathsf{fst}}
\newcommand{\Fst}[1]{\FstSym(#1)}
\newcommand{\SndSym}{\mathsf{snd}}
\newcommand{\Snd}[1]{\SndSym(#1)}

\newcommand{\DomainSym}{\mathsf{domain}}
\newcommand{\Domain}[1]{\DomainSym(#1)}
\newcommand{\RangeSym}{\mathsf{range}}
\newcommand{\Range}[1]{\RangeSym(#1)}
\newcommand{\CompatibleClosure}[2][]{\Bracks{#2}^{#1}}

\newcommand{\GroundRelation}[2]{\mathrel{{\sarrow{->}{#1}}_{#2}}}
\newcommand{\ReflTransRelation}[2]{\mathrel{{\sarrow{->>}{#1}}_{#2}}}
\newcommand{\ReflTransSymmRelation}[2]{\mathrel{{\sarrow{<<->>}{#1}}_{#2}}}

\newcommand{\CompatibleRelation}[3][]{\mathrel{{\sarrow{->}{\CompatibleClosure[#1]{#2}}}_{#3}}}

\renewcommand{\PerturbFunAt}[3]{\Ternary{}{#1}{\lbrack}{#2}{\mathbin{\mapsto}}{#3}{\rbrack}}

\newcommand{\MetaSubst}[3]{\TernaryB{}{#1}{[}{#2}{\mathbin{\coloneqq}\relax}{#3}{]}}

\newcommand{\EquivClass}[2]{[{#1}]_{#2}}

\newcommand{\Hole}{}          
\newcommand{\HoleW}{\ContextHole} 

\newcommand{\IParens}[1]{(#1)}
\newcommand{\INullary}[1]{#1}
\newcommand{\IUnary}[3]{#1#2#3}
\newcommand{\IBinary}[5]{#1#2#3#4#5}
\newcommand{\IBinaryX}[5]{#1#2^{#3#4#5}}
\newcommand{\ITernary}[7]{#1#2#3#4#5#6#7}

\newcommand{\TParens}[1]{#1}
\newcommand{\TNullary}[1]{\TR{\Inline #1}}
\newcommand{\TUnary}[3]{\pstree{\TR{\Inline #1{\Hole}#3}}{{#2}}}
\newcommand{\TUnaryW}[3]{\pstree{\TR{\Inline #1\HoleW#3}}{{#2}}}
\newcommand{\TBinary}[5]{\pstree{\TR{\Inline #1\Hole#3\Hole#5}}{{#2}{#4}}}
\newcommand{\TBinaryW}[5]{\pstree{\TR{\Inline #1\HoleW#3\HoleW#5}}{{#2}{#4}}}
\newcommand{\TBinaryX}[5]{\TR{\Inline #1#2^{#3#4#5}}}
\newcommand{\TTernary}[7]{\pstree{\TR{\Inline #1\Hole#3\Hole#5\Hole#7}}{{#2}{#4}{#6}}}
\newcommand{\TTernaryA}[7]{\pstree{\TR{\Inline #1#2#3\HoleW#5\HoleW#7}}{{#4}{#6}}}
\newcommand{\TTernaryB}[7]{\pstree{\TR{\Inline #1\Hole#3#4#5\Hole#7}}{{#2}{#6}}}

\newcommand{\Inline}{
  \let\Parens=\IParens
  \let\Nullary=\INullary
  \let\Unary=\IUnary
  \let\UnaryW=\IUnary
  \let\Binary=\IBinary
  \let\BinaryW=\IBinary
  \let\BinaryX=\IBinaryX
  \let\Ternary=\ITernary
  \let\TernaryA=\ITernary
  \let\TernaryB=\ITernary
}
\newcommand{\JTree}{
  \let\Parens=\TParens
  \let\Nullary=\TNullary
  \let\Unary=\TUnary
  \let\UnaryW=\TUnaryW
  \let\Binary=\TBinary
  \let\BinaryW=\TBinaryW
  \let\BinaryX=\TBinaryX
  \let\Ternary=\TTernary
  \let\TernaryA=\TTernaryA
  \let\TernaryB=\TTernaryB
}

\Inline

\newcommand{\NatSet}{\mathbb{N}}
\newcommand{\PointerSet}{\mtnormsf{Pointer}}
\newcommand{\PositionSet}{\mtnormsf{Pos}}
\newcommand{\BarSign}{\overline{\mtnormsf{B}}}
\newcommand{\CoreSet}{\mtnormsf{Core}}

\newcommand{\ContextSet}[1]{{#1}\mtnormsf{-Context}}

\newcommand{\TyVarSet}{\mtnormsf{Ty-Variable}}
\newcommand{\SimpleTypeSet}{\mtnormsf{Simple-Type}}

\newcommand{\AnyX}{\metavar{X}}
\newcommand{\AnyY}{\metavar{Y}}
\newcommand{\AnyZ}{\metavar{Z}}

\newcommand{\MetaVar}{\nu}

\newcommand{\SetS}{\mathcal{S}}
\newcommand{\SetT}{\mathcal{T}}
\newcommand{\RelationR}{\mathcal{R}}

\newcommand{\metavar}[1]{\Nullary{#1}}

\newcommand{\TermVarX}{\metavar{x}}

\newcommand{\TyVarA}{\metavar{a}}
\newcommand{\TyVarB}{\metavar{b}}

\newcommand{\Constructor}{\metavar{c}}

\newcommand{\Alternative}{\mathcal{A}}
\newcommand{\Type}{\metavar{T}}

\newcommand{\Object}{\metavar{O}}
\newcommand{\ObjectSet}{\mtnormsf{Object}}
\newcommand{\Arrangement}{\metavar{A}}
\newcommand{\CoreArrangement}{\metavar{\hat{A}}}
\newcommand{\ArrangementSet}{\mtnormsf{Arrangement}}
\newcommand{\ArrangementEquiv}{\approx}
\newcommand{\ArrangementEquivClass}[1]{\EquivClass{#1}{\ArrangementEquiv}}
\newcommand{\Symbol}{\metavar{s}}
\newcommand{\SymbolSet}{\mtnormsf{Symbol}}
\newcommand{\FillContextAux}[2]{\mtnormsf{fill}\Parens{{#1},{#2}}}
\newcommand{\SwapNamesIn}[3]{\mtnormsf{swap}\Parens{{#1},{#2},{#3}}}
\newcommand{\AlphaConvertibleRel}{\equiv_{\upalpha}}
\newcommand{\NamesInSameGroupRel}{\sim}

\newcommand{\SubstSym}{\coloneqq}
\newcommand{\Subst}[3]{\Ternary{}{#1}{\mathbin{\SubstSym}}{#2}{,}{#3}{}}

\newcommand{\ConcreteTermVar}[1]{\mathsf{x}_{#1}}

\newcommand{\Abs}[2]{\Binary{\upsflambda}{#1}{.}{#2}{}}
\newcommand{\AbsP}[2]{\Parens{\Abs{#1}{#2}}}

\newcommand{\App}[2]{\Binary{}{#1}{\mathbin{\AppSym}}{#2}{}}
\newcommand{\AppP}[2]{\Parens{\App{#1}{#2}}}

\newcommand{\LetIn}[3]{\mathsf{let}\ #1\ {=}\ #2\ \mathsf{in}\ #3}
\newcommand{\LetInP}[3]{\Parens{\LetIn{#1}{#2}{#3}}}

\newcommand{\ConcreteTyVar}[1]{\mathsf{a}_{#1}}

\newcommand{\FunType}[2]{\Binary{}{#1}{\mathbin{\FunTypeSym}}{#2}{}}

\newcommand{\FunTypeP}[2]{\Parens{\FunType{#1}{#2}}}

\newcommand{\TupleLSym}{(}
\newcommand{\TupleRSym}{)}
\newcommand{\Pair}[2]{\Binary{\TupleLSym}{#1}{,}{#2}{\TupleRSym}}
\newcommand{\Triple}[3]{\Ternary{\TupleLSym}{#1}{,}{#2}{,}{#3}{\TupleRSym}}
\newcommand{\Tuple}{\DelimWrapper{\TupleLSym}{\TupleRSym}}

\newcommand{\SeqLSym}{[}
\newcommand{\SeqRSym}{]}
\newcommand{\SeqVar}[1]{\vec{#1}}
\newcommand{\SeqSet}[1]{{#1}^*}
\renewcommand{\Seq}{\DelimWrapper{\SeqLSym}{\SeqRSym}}
\renewcommand{\SeqEmpty}{\Seq{}}
\renewcommand{\SeqLen}[1]{\Verts{#1}}
\newcommand{\SeqConcatSym}{\cdot}
\renewcommand{\SeqConcat}[2]{\Binary{}{#1}{\SeqConcatSym}{#2}{}}
\newcommand{\SeqConcatP}[2]{\Parens{\SeqConcat{#1}{#2}}}
\newcommand{\Nth}[2]{#1(#2)}

\newcommand{\ContextHole}{\Box}

\newcommand{\FillContext}[2]{{#1}\Seq{#2}}

\newcommand{\Mark}[1]{\underline{\vrule width 0em depth 0.5ex height 0ex #1}}
\newcommand{\FreeVars}[1]{\mtnormsf{FN}(#1)}

\notetrue
\notefalse

\ifjbw
\else
\fi

\usepackage{qtree}

\usepackage{mathtools}
\usepackage{graphicx}

\makeatletter
\newcommand{\myLines}[1]{
\begin{picture}(1,1)
\put(0,0){\vector(0,1){1}}
\end{picture}}
\let\qdrawReal=\qdraw@branches
\newcommand\brOverride{\let\qdraw@branches=\myLines}
\newcommand\brRestore{\let\qdraw@branches=\qdrawReal}
\makeatother

\title{What Does This Notation Mean Anyway?\\ 
\begin{large} 
  BNF-Style Notation as it is Actually Used
\end{large}}

\author{D.~A.~Feller \and J.~B.~Wells \and S\'ebastien Carlier \and F. Kamareddine}

\def\titlerunning{What Does This Notation Mean Anyway?}
\def\authorrunning{D.~A.~Feller, J.~B.~Wells, F. Kamareddine, S. Carlier}

\providecommand{\event}{LFMTP 2018}

\begin{document}

\maketitle




\notefalse
\note{}

\begin{abstract}
Following the introduction of BNF notation by Backus for the Algol 60 report and subsequent notational variants, a metalanguage involving formal
“grammars” has developed for discussing structured objects in Computer Science and Mathematical 
Logic.\note{}
We refer to this offspring of BNF as \emph{Math-BNF} or \emph{MBNF},
to the original BNF and its notational variants just as \emph{BNF},
and to aspects common to both as \emph{BNF-style}.
What all BNF-style notations share is the use of production rules roughly of this form:
$$
  \bullet \mathrel{::=} \circ_1 \mid \cdots \mid \circ_n
$$
Normally, such a rule says ``every instance of $\circ_i$ for $i \in \{1, \ldots, n\}$ is also an instance of $\bullet$''.

MBNF is distinct from BNF in the entities and operations it allows.
Instead of strings, MBNF builds arrangements of symbols that we call
\emph{math-text}\note{}.
Sometimes ``syntax'' is defined by interleaving MBNF production rules
and other mathematical definitions that can contain chunks of
math-text.

There is no clear definition of MBNF.
Readers do not have a document which tells them how MBNF is to be read
and must learn MBNF through a process of cultural initiation.
To the extent that MBNF is defined, it is largely through examples
scattered throughout the literature\longonly{ and which require
  readers to guess the mathematical structures underpinning them}.

This paper gives MBNF examples illustrating some of the differences
between MBNF and BNF.
We propose a definition of \emph{syntactic math text} (SMT) which
handles many (but far from all) uses of math-text and MBNF in the
wild.
We aim to balance the goal of being accessible and not requiring too
much prerequisite knowledge with the conflicting goal of providing a
rich mathematical structure that already supports many uses and has
possibilities to be extended to support more challenging cases.
\end{abstract}

\section{Background and Motivation}

Understanding MBNF is important to interpreting papers in theoretical
computer science.
Out of the 30 papers in the ESOP 2012 proceedings~\cite{esop2012}, 19
used MBNF, while not one used BNF.\footnote{We chose ESOP 2012 because
  its book was the most recent conference proceedings that we had as a paper book.
  Because the first book we picked contained an abundance of
  challenging instances of MBNF, our wider searching has mainly been
  to find even more challenging examples.
  We will be happy to receive pointers to additional interesting cases.
  \note{}}
This section highlights some of the ways in which the notation we call
MBNF differs from BNF.
This should demonstrate that a definition could be helpful.

\iflong
  \let\subsectionwhenlong=\subsection
\else
  \let\subsectionwhenlong=\paragraph
\fi

\subsectionwhenlong{Where BNF uses Strings, MBNF Uses Math-Text}

In addition to arranging symbols from left to right on the page,
math-text allows subscripting, superscripting,\note{} and placing text above or below other text.
It also allows for marking whole segments of text, for example with an
overbar (a vinculum).
Readers can find more detailed information on how math-text can be laid out in The TeXbook \cite{TeXbook}, or the Presentation MathML \cite{Ion:01:MML} and OpenDocument \cite{ISO26300} standards.
Here is a nonsense piece of Math-text to illustrate how it may be laid out:
$${}^c\!\!\downarrow a'= \check{p}\langle v''_x \odot a^{2+1}\rangle- \overline{f_x^n +\overline{y\cdot fj}}+\sum\limits_{i=0}^\infty s_{i\in 1\ldots n}\sarrow{->}{a,b,c} {b}\hat{a}$$

Instead of non-terminal symbols, MBNF uses
\emph{metavariables}\footnote{We use metavariable to mean a variable
  at the meta-level which denotes something at an object-level.},
which appear in math-text and obey the conventions of mathematical
variables.
Metavariables are not distinguished from other symbols by annotating
them as BNF does, but by font, spacing, or merely tradition.

Parentheses for disambiguation are not needed in MBNF grammars and
when an MBNF grammar specifies such parentheses they can often be
omitted without any need to explain.
When possible, MBNF takes advantage of the tree-like structure
implicit in the layout of symbols on the page when features like
superscripting and overbarring are used.

\subsectionwhenlong{MBNF Is Aimed at Human Readers}

MBNF is meant to be interpreted by humans, not computers/parser
generators.
\note{}%
It is common to define a MBNF grammar in an article for humans and a
separate EBNF grammar for use with a parser generator to build a
corresponding implementation.
Entities defined with MBNF are not intended or expected to be
serialized or parsed and MBNF grammars are typically missing features
needed to disambiguate complex terms.
Papers often put complicated uses of the mathematical metalanguage in
the middle of MBNF notation.

\note{}%
\note{}%

\subsectionwhenlong{MBNF Allows Powerful Operators Like Context Hole Filling (a.k.a.\ Tree Splicing)}

Chang and Felleisen~\cite[p 134]{esopcallneed} present an MBNF grammar
defining the $\lambda$-term contexts with one hole where the
spine\footnote{The root node is on the spine.
If $A$ is applied to $B$ by an application on the spine, the root node of $A$ is on the spine and the root node of $B$ is not.
  If a node on the spine is an abstraction each of its children is on
  the spine.} is a balanced segment\footnote{A balanced segment is one
  where each application has a matching abstraction and where each
  application/abstraction pair contains a balanced segment.} ending in a hole.
For explanatory purposes we alter their grammar slightly by writing
${\color{blue} e}{\color{\ColorA} @}{\color{blue} e}$ instead of
${\color{blue} e}\,{\color{blue} e}$ and adding parentheses.
{{\color{\ColorA} Concrete syntax} \note{}and
  {\color{\ColorA} BNF-style notation} \note{}are
  {\color{\ColorA} green}.
{\color{blue} Metavariables} \note{}are
{\color{blue} blue}.
{\color{\ColorB} Additional operators}
\note{}are {\color{\ColorB} red}.
$$
 \arraycolsep=2pt
 \begin{array}{rcl}
 {\color{blue} e}&\mathrel{\color{\ColorA} ::=}&{\color{blue} x}\mathbin{\color{\ColorA} \mid}{\color{\ColorA} (}{\color{\ColorA} \lambda}
  {\color{blue} x}{\color{\ColorA} .}{\color{blue} e}{\color{\ColorA} )}\mathbin{\color{\ColorA} \mid}{\color{\ColorA} (}{\color{blue} e}
  {\color{\ColorA} @}{\color{blue} e}{\color{\ColorA} )}\\
 {\color{blue} A}&\mathrel{\color{\ColorA} ::=}&{\color{\ColorA} [\, ]}\mathbin{\color{\ColorA} \mid }{\color{\ColorA} (}{\color{blue} A}
 {\color{\ColorB} [}{\color{\ColorA} (}{\color{\ColorA} \lambda} {\color{blue} x}{\color{\ColorA} .}{\color{blue} A}{\color{\ColorA} )}
 {\color{\ColorB} ]}{\color{\ColorA} @}{\color{blue} e}{\color{\ColorA} )}
 \end{array}
$$
One can think of the context hole filling operation in this grammar
({\color{red}$[\, ]$} in
${\color{\ColorA} (}{\color{blue} A}{\color{\ColorB} [}{\color{\ColorA} (}{\color{\ColorA} \lambda} {\color{blue} x}{\color{\ColorA} .}{\color{blue} A}{\color{\ColorA} )}{\color{\ColorB} ]}{\color{\ColorA} @}{\color{blue} e}{\color{\ColorA} )}$) 
  as performing tree splicing operations within the syntax.
Consider these trees which illustrate steps in building syntax trees for ${\color{blue}A}$:

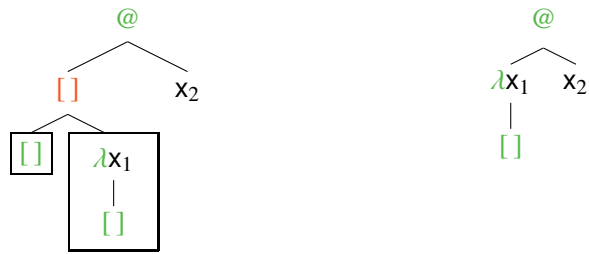
\Tree [.{\color{\ColorA}$@$} [.{\color{red}$[\, ]$} \framebox{{\color{\ColorA}$[\, ]$}} [.{\color{\ColorA}$\lambda$}$\ConcreteTermVar{1}$ {\color{\ColorA}$[\, ]$}  ] !{\qframesubtree} ] $\ConcreteTermVar{2}$ ]
\Tree [.{\color{\ColorA}$@$}  [.{\color{\ColorA}$\lambda$}$\ConcreteTermVar{1}$   {\color{\ColorA}$[\, ]$} ]   $\ConcreteTermVar{2}$ ] 
\\

These trees show the result of the second rule  where each ${\color{blue} A}$ is {\color{\ColorA}$[\, ]$} and {\color{blue}$e$} is a variable.
The tree on the left is the tree corresponding to 
${\color{blue} A}{\color{\ColorB} [}{\color{\ColorA} \lambda} {\color{blue} x}{\color{\ColorA} .}{\color{blue} A}{\color{\ColorB} ]}{\color{\ColorA} @}{\color{blue} e}$
 before the hole filling operation is performed, where the first ${\color{blue} A}$ is assigned {\color{\ColorA}$[\, ]$}. 
The tree on the right represents an unparsing of what we would normally consider the syntax tree for {\color{\ColorA}((}{\color{\ColorA}$\lambda$}$x_1${\color{\ColorA}.$[\, ]$}{\color{\ColorA})}{\color{\ColorA} $@$}$x_2${\color{\ColorA})}.
We write $\ConcreteTermVar{1}$ and $\ConcreteTermVar{2}$ for disambiguated instances of {\color{blue}$x$}.
A metavariable assigned a value won't appear in the final tree.
If it's not a terminal node, {\color{red}$[\, ]$} tells us to fill in the leaf in the frame on the left with the the tree in the frame on the right.
Once performed, {\color{red}$[\, ]$} disappears.

\longonly{We can show that }Unlike BNF, the ``language'' of the
metavariable/non-terminal $\color{blue}A$ (the set of strings derived
from $\color{blue}A$ using roughly the rules of BNF plus hole filling)
is not context-free and so MBNF certainly
isn't.\note{
}

\subsectionwhenlong{MBNF Mixes Math Stuff With BNF-Style Notation}
Germane and Might \cite[pg 20]{Germane:2017:PEA:3009837.3009899} mix BNF-style notation freely with mathematical notation in such a way that the resulting grammar relies upon both sets produced from the result of MBNF calculations and MBNF production rules which use metavariables defined using mathematical notation:
  $$
  \arraycolsep=2pt
 \begin{array}{rcl@{\qquad}rcl}
 u\in \mathit{UVar} &=& \mathtt{a\, set\, of\, identifiers}\,\,&
 \mathit{ccall}\in   \mathit{CCall} &\mathrel{::=}& (q\, e^*)_\gamma  \\
k\in   \mathit{CVar} &=& \mathtt{a\, set\, of\, identifiers}\,& e, f\in   \mathit{UExp} &=& \mathit{UVar}+\mathit{ULam}\\
   \mathit{lam}\in   \mathit{Lam} &=& \mathit{ULam}+\mathit{CLam}\quad &q\in   \mathit{CExp} &=& \mathit{CVar}+\mathit{CLam}\\
   \mathit{ulam}\in   \mathit{ULam} &\mathrel{::=}& (\lambda e(u^*k)call)\quad & 
   \ell\in   \mathit{ULab}&=&\mathtt{a\, set\, of\, labels} \\
   \mathit{clam}\in   \mathit{CLam} &\mathrel{::=}& (\lambda_\gamma (u^*)call)&
   \gamma\in  \mathit{CLab}&=&\mathtt{a\, set\, of\, labels}\\
   \mathit{call}\in   \mathit{Call} &=& \mathit{UCall}+\mathit{CCall}& 
   \\
   \mathit{ucall}\in   \mathit{UCall} &\mathrel{::=}& (f e^*q)_\ell& &  & 
   \end{array}
   $$\note{}
   The results of math computations are interleaved with MBNF production rules, not just applied after the results of the production rules have been obtained.
   This grammar uses $\bullet_1\in\bullet_2$ to mean ``$\bullet_2$ is the language of $\bullet_1$" (this is the case in both the MBNF production rules ($::=$) and the math itself ($=$)).

   \note{}
    
\note{}

\subsectionwhenlong{MBNF Has at Least the Power of Indexed Grammars}

Inoe and Taha~\cite[pg 361]{multistage} use this MBNF:
$$
\mathcal{E}^{\ell ,m}\in \mathit{ECtx}^{\ell ,m}_\mathbf{n}\mathrel{::=}\cdots\mid \langle\mathcal{E}^{\ell +1,m}\rangle\mid \cdots
$$
This suggests that MBNF deals with the family of indexed grammars \cite[p 389-390]{indexed}, which is yet another reason it's not context-free.
The $\ell +1$ is a calculation that is not intended to be part of the syntax.
The production rule above defines an infinite set of metavariables
ranging over different sets.

\subsectionwhenlong{MBNF Allows Arbitrary Side Conditions on Production Rules}

An example of a production rule with a side condition can be found in Chang and Felleisen \cite[p 134]{esopcallneed}:
$$
  E= [\text{ }]\mid Ee\mid A[E]\mid \hat{A}[A[\lambda x.\check{A}[E[x]]]E]\qquad\qquad\text{ where }\hat{A}[\check{A}]\in A
$$
\note{}%
It is possible to make side conditions that prevent MBNF rules from
having a solution.
A definition for MBNF can help in finding restrictions on side conditions
that ensure MBNF rules actually define something.
\note{
}

\subsectionwhenlong{MBNF ``Syntax" Can Contain Very Large Infinite Sets}

Toronto and McCarthy \cite[p 297]{lambdazfc} use the following MBNF:
$$e::=\cdots\mid\langle t_\mathit{set},\{ e^{*\kappa}\}\rangle$$
We are told $\{ e^{*\kappa}\}$ denotes ``sets comprised of no more
than $\kappa$ terms from the language of $e$''.
The author does not state what $\kappa$ is, but elsewhere in the paper it is an inaccessible cardinal. 
It seems as though $\kappa$ is also intended to be an inaccessible cardinal here.
This section of an MBNF for $e$ is taken from a larger MBNF that contains a term which ranges over all the encodings of all the hereditarily accessible sets.
BNF, by contrast, only deals with strings of finite length.

\subsectionwhenlong{MBNF Allows Infinitary Operators}
Fdo, D\'{i}az and N\'{u}\~{n}ez \cite[p 539]{inftyop} write an MBNF with the following operator:
$$P::=\cdots\mid\bigsqcap\limits_{i\in I}P_i\mid\cdots$$
The authors state this is infinitary (i.e. we should regard $I$ to be infinite). 
The authors tell us the MBNF this is taken from is defined by regarding (M)BNF expressions as fixed point equations and a least fixed point can be found by bounding the size of the possible set of indices by some infinite cardinal.

We may think of infinitary operators as allowing us to define trees of infinite breadth (i.e. trees whose internal nodes may have infinitely many direct children), where BNF only deals with   finite strings.

\subsectionwhenlong{MBNF Allows Co-Inductive Definitions}
Eberhart, Hirschowitz and Seiller \cite[p 94]{eberhart_et_al:LIPIcs:2015:5528} intend the following MBNF to define infinite terms co-inductively:

$$\arraycolsep=2pt
\begin{array}{rcl}
P,Q&\mathrel{::=}&\Sigma_{i\in n}G_i \text{ }{\large\mid }\text{ } (P{\small\lvert }Q)\\
G&\mathrel{::=}&\overline{a}\langle b\rangle .P\text{ } {\large\mid } \text{ }a(b).P\text{ } {\large\mid } \text{ }\nu a.P\text{ } {\large\mid } \text{ }\tau .P\text{ } {\large\mid } \text{ }\heartsuit .P
\end{array}$$

\note{}

We may think of co-inductive definitions as allowing us to define trees of infinite depth (i.e. trees in which paths
 may pass through infinitely many nodes), where BNF only deals with  finite strings.
\note{}

\section{A Method to Allow Reading Some Uses of Mathematical ``Syntax"}
\label{sec:syntax-modulo-equivalences}

This section defines \emph{syntactic math text} (SMT) which will allow
reading some uses of math text as being  ``syntax" and standing for
essentially themselves, e.g., $1+3$ can continue to stand for $4$
while $\lambda x.x$ can in some sense stand for itself.
SMT plus a definition of the $::=$ notation allows us to interpret the more common uses of MBNF as they are written.
It also provides some support for more complicated uses with a little extra machinery.
We do not aim to cover every use of MBNF in the literature, but we hope to provide a good foundation which can be built upon.

As well as dealing with some of MBNF, SMT provides a more general notion of \emph{objects} appearing within syntax that behave like equivalences over chunks of math-text representing syntax.
This enables us to interpret working modulo equivalences on math-text representing syntax.

Kamareddine et al.~\cite{computerising} make the point that converting mathematical text to a form where it can be checked by a proof assistant is a process that involves both human input and intermediary translations.
Our proposal focuses on the translation, performed by the reader, of
math-text used to define syntax, as it appears in a document, to a
more formal structure, which is not encoded in the language of a
theorem prover or proof assistant.

Our proposal relies as much as possible on the mathematical meta-level.
For example, we use ellipses and related methods for abbreviating
sequences from the mathematical meta-level.
    Incomplete definitions (relying on some choice of metavariable) cause the resulting grammar to be defined as the output of a function depending on this choice.
Any otherwise pointless statement of the form $x \in S$ declares $x$
and any \emph{decorated} $x$ (e.g., $x_1$, $x_2$, $\ldots$, $x'$,
$x''$, etc.)\ as a variable ranging over $S$.
\note{}
    
Our proposal is intended to be descriptive rather than prescriptive.
    We aim to handle both historical documents and new works.
For published uses of MBNF that our proposal fails to handle, this is
a problem to be solved in future work.
We do not aim at displacing the input languages of proof assistants or
syntactic variants of BNF which already have solid definitions.

\note{}

\note{
}

\note{
}

\subsection{Objects, Arrangements, and Symbols}
\label{sec:objects-arrangements-symbols}

\note{ 
  }

We now define the main notion of syntactic \DefiningUse{objects} and
the auxiliary notion of \DefiningUse{arrangements}.
In essence, syntactic objects are arrangements of \DefiningUse{symbols}, numbers,
and pointers to subobjects, where the arrangement can include left-to-right
sequencing, superscripting, subscripting\note{} and 
overlining.\note{}
  We use pointers to subobjects inside objects rather than the subobjects themselves, because the sets within the model for objects would be too large otherwise and because we wanted to allow for objects to be nested within themselves, provided some syntax is added as part of this nesting.
To support $\upalpha$-conversion and operators that are associative,
commutative, idempotent, etc., the objects are defined so that in
effect they work modulo an equivalence relation on arrangements that is
defined separately.

\note{}
Let $\Symbol$ range over the set $\SymbolSet$ containing syntactic
\DefiningUse{symbols} to be used in arrangements.
We require that $\SymbolSet$ is disjoint from all other sets defined
here.
We also require that some symbols are \emph{not} in $\SymbolSet$, namely
the square brackets (``['' and ``]'') and the special square symbol
$\ContextHole$ (which represents a hole in which an object can be
placed).
The symbols can include letters, parentheses and other
parenthesis-like symbols (e.g., $\bbblangle$ and $\bbbrangle$ and
$\llbrack$ and $\rrbrack$), punctuation, and other symbols.
Letters (Roman or Greek) used as syntactic symbols will be typeset
using an upright sans-serif font to distinguish them from
metavariables which are written in a slanted serif font (generally
italics).
For example, $\mathsf{a}$, $\mathsf{C}$, $\upsflambda$, and
$\mathsf{\Gamma}$ could be syntactic symbols while $a$, $C$,
$\lambda$, and $\mathit{\Gamma}$ would be metavariables.
We avoid using any particular letter both ways, except for symbols
used in \FirstUse{names}, where for example $\ConcreteTermVar{i}$
could be a syntactic name at the same time as $\TermVarX$ could be a
metavariable ranging over names (see
\autoref{sec:names-binding-alpha-conversion-substitution}\note{}).

\note{}

The set $\ObjectSet$ of syntactic \DefiningUse{objects} and the set
$\ArrangementSet$ of syntactic \DefiningUse{arrangements} are defined
simultaneously.
Let $\Object$ range over $\ObjectSet$ and let $\Arrangement$ range
over $\ArrangementSet$.
We represent each object by a member of the set $\PointerSet$.
Let $P_\Object$ be the pointer that indicates $\Object$.
Let ${\ArrangementEquiv}\subset{\ArrangementSet\times\ArrangementSet}$
be an equivalence relation that is reflexive on $\ArrangementSet$.
We require that if $\Arrangement_1\ArrangementEquiv\Arrangement_2$ and
$\Arrangement_1\ne\Arrangement_2$, then neither $\Arrangement_1$ nor
$\Arrangement_2$ may have used the special object $\ContextHole$ in
their construction.

The sets $\ObjectSet$ and $\ArrangementSet$ are the smallest sets
satisfying the following conditions.
\begin{enumerate}
\item
  The \DefiningUse{empty arrangement} $\epsilon$ is in $\ArrangementSet$.
\item
  The core items of arrangements are symbols, pointers to objects, numbers, and
 \note{} overlined arrangements.
  For any symbol $\Symbol$, pointer $P_\Object$, number $n\in\NatSet$,
  and nonempty arrangement $\Arrangement\neq\epsilon$, all of the following are in
  $\ArrangementSet$: $\Symbol$, $P_\Object$, $n$,
  \note{} and $\overline{\Arrangement}$.
  Furthermore, these are all \DefiningUse{core arrangements}, which
  are ranged over by the metavariable $\CoreArrangement$.
\item
  Left-to-right sequencing allows appending additional core
  arrangements to a non-empt arrangement.
  For any arrangement $\Arrangement\neq\epsilon$ and core arrangement
  $\CoreArrangement$, it holds that $\Arrangement\CoreArrangement$ is
  in $\ArrangementSet$.
\item
  Superscripting, subscripting\note{} etc.\ are supported.
  For non-empty arrangements $\Arrangement$, $\Arrangement_1$ and
  $\Arrangement_2$, all of the following are in $\ArrangementSet$:
  $\Arrangement^{\Arrangement_1}$, $\Arrangement_{\Arrangement_2}$,
  $\Arrangement^{\Arrangement_1}_{\Arrangement_2}$, ${}^{\Arrangement_1}\Arrangement$...

\item
  If $\SetS$ contains does not contain any arrangements consisting of a bare pointer to an object, $\SetS$ is non-empty, and $|\SetS |\leq\aleph_0$, then $\SetS\in\ObjectSet$.

  If $\SetS\subset\ArrangementSet$ is not an equivalence class of
  $\ArrangementEquiv$ or any
  members of $\SetS$ are ill-formed, then $\SetS$ is
  \DefiningUse{ill-formed}.
  An arrangement is ill-formed iff any of its subcomponents is
  ill-formed.  (Symbols and natural numbers are well formed.)

  (Thus, it is allowed to build an object from ill-formed
  arrangements, and the resulting object is ill-formed.)
\item
    There is a special symbol in $\ObjectSet$ indicating a \DefiningUse{hole} $\ContextHole$ in which an object is to
  be placed.
  
 \note{}
\end{enumerate}

\note{}
There are various reasons why we have built equivalence classes into
arrangements rather than making them identical to math-text.
   We want to eventually support math stuff in syntax, with math stuff
    containing objects not arrangements.
    We want to allow object-to-object operations in production rules.
    When we define equivalences inductively over arrangements we want some of that structure to be represented by our model.

We write $\ArrangementEquivClass{\Arrangement}$ for the object that
contains all the arrangements equivalent to $\Arrangement$ by the
equivalence relation $\ArrangementEquiv$.
Only objects of the form  $\ArrangementEquivClass{\Arrangement}$ are \emph{well formed}.

\subsection{Syntax Shorthand: Arrangement Coercions}
From \autoref{ex:objects}, the reader will observe that it's cumbersome to write $P_\Object$ in so many places when all we're interested in is the identity for objects.
We introduce the following convention:
\begin{convention}[Coercing Objects to Pointers]
\label{conv:implicit pointers}
We allow $\Object$ to be written instead of $P_\Object$ in an arrangement.
\end{convention}

\begin{example}
The expression $\ArrangementEquivClass{\lambda\Object_1 .\Object_2}$, stands for
 $ \ArrangementEquivClass{\lambda P_{\Object_1}.P_{\Object_2}}$.
\end{example}

\label{sec:syntax-shorthand-implicit-coercions}
We define \emph{meta-level parentheses} to be those parentheses which surround a single object and which may optionally be omitted  from some arrangements with a similar form.\footnote{We largely leave it up to the reader to determine which parentheses are meta-level.
If a primitive constructor (section 2.4) appears inside some arrangements with parentheses surrounding it and other arrangements without them, it usually indicates these parentheses are meta-level.
Similarly, parentheses which only surround a single metavariable corresponding to an object are frequently meta-level.
Parentheses surrounding syntax which is to be thought of as a sequence are normally not meta-level.
To help with this ambiguity, from this point forward all parentheses appearing in arrangements inside this document are meta-level.}

It
is still cumbersome to write $\ArrangementEquivClass{\:\cdot\:}$ in so
many places.
One of the ways we deal with this is to arrange for this to happen
automatically at places where a piece of meta-level syntax requires an arrangement to be regarded as an object.

\begin{convention}[Coercing Arrangements to Objects]
  \label{conv:arrangement-object-coercion}
  We require that when an arrangement $\Arrangement$ is written, but
  the surrounding context only makes sense if the value of the
  expression is an object, then the arrangement $\Arrangement$ is
  implicitly coerced to the object
  $\ArrangementEquivClass{\Arrangement}$, as though the latter
  had been written instead.
  As a special case of this, we require that an arrangement that containing meta-level parentheses is to be read as though the parentheses
  were instead a use of $\ArrangementEquivClass{\:\cdot\:}$.
\end{convention}

\begin{convention}[Coercing Arrangements to Pointers]
  \label{conv:arrangement-pointer-coercion}
  We require that when an arrangement $\Arrangement$ is written, but
  the surrounding context only makes sense if the value of the
  expression is a pointer, then the arrangement $\Arrangement$ is implicitly coerced to the pointer to the object given by \autoref{conv:arrangement-object-coercion}.
\end{convention}

Due to the combination of \autoref{conv:arrangement-object-coercion},  \autoref{conv:arrangement-pointer-coercion}
and the tight restrictions on where round parentheses can occur in
proper arrangements, most uses of round parentheses will not be
symbols that are part of syntactic arrangements but instead will be
part of the meta-level mathematical reasoning.

\begin{example}
  The expression $(\Object_1\,{\Object_2})\,{\Object_3}$, which contains meta-level parentheses,
    stands for
  $\ArrangementEquivClass{{\Object_1}\,{\Object_2}}\,{\Object_3}$.
  \note{}
  If we write $\Object={({\Object_1}\,{\Object_2})}\,{\Object_3}$,
  then this stands for writing
  $\Object=\ArrangementEquivClass{{\ArrangementEquivClass{{\Object_1}\,{\Object_2}}}\,{\Object_3}}$,
  because the equation's left-hand side must be an object due to the
  declaration that the metavariable $\Object$ ranges over
  $\ObjectSet$.
 \note{}
\end{example}

We have left $\ArrangementEquiv$ mostly unspecified so far.
The sets $\ObjectSet$ and $\ArrangementSet$ do not depend on
$\ArrangementEquiv$, but their subsets of well formed objects and
arrangements do depend on $\ArrangementEquiv$.
The definition of $\ArrangementEquiv$ may be adjusted by the authors of a paper at any point, and the set of well formed objects in scope
 will therefore change at the times these adjustments are
made.
The effect of \autoref{conv:arrangement-object-coercion} will
similarly change; the same expression can denote different
objects at different places if there is an
intervening change to  $\ArrangementEquiv$.

\subsection{Contexts and Hole Filling}

A \DefiningUse{context} is an object $\Object\in\ObjectSet$ with at
least one use of the special hole object $\ContextHole$.
The number of hole symbols in an object or arrangement is its \DefiningUse{arity}.
We now define \DefiningUse{context-hole filling} for arbitrary objects
and arrangements (although it will in general only do something useful for well formed
objects and arrangements with the correct arity).
Let the operations $\Object\Seq{\Object_1,\ldots,\Object_n}$, $P_{\Object}\Seq{\Object_1,\ldots,\Object_n}$  and
$\Arrangement\Seq{\Object_1,\ldots,\Object_n}$ which fill the holes reachable from
$\Object$, $P_\Object$ and $\Arrangement$ with the objects in the sequence
$\SeqVar{\Object}=\Seq{\Object_1,\ldots,\Object_n}$ be defined as
follows:

\begin{enumerate}
\item
 $P_\Object\SeqVar{\Object}=P_{\Object'}$  and $\Object\SeqVar{\Object}=\Object'$ iff
  $\FillContextAux{\Object}{\SeqVar{\Object}}=\Pair{\Object'}{\SeqEmpty}$.
  Similarly, $\Arrangement\SeqVar{\Object}=\Arrangement'$ iff
  $\FillContextAux{\Arrangement}{\SeqVar{\Object}}=\Pair{\Arrangement'}{\SeqEmpty}$.
  The results of $\FillContextAux{\Object}{\SeqVar{\Object}}$ and
  $\FillContextAux{\Arrangement}{\SeqVar{\Object}}$ are undefined
  except where explicitly defined below.
  (The result is undefined unless all of the replacements
  are used, so the number of replacements must match the arity.)
\item
  \(
      \FillContextAux{\ContextHole}{\SeqConcat{\Seq{\Object}}{\SeqVar{\Object}}}
    = \Pair{\Object}{\SeqVar{\Object}}
  \).
  (Each hole uses up one of the replacements.)
\item
  \(
      \FillContextAux{\Set{\Arrangement}}{\SeqVar{\Object}}
    = \Pair{\ArrangementEquivClass{\Arrangement'}}{\SeqVar{\Object}'}
  \)
  if
  \(
      \FillContextAux{\Arrangement}{\SeqVar{\Object}}
    = \Pair{\Arrangement'}{\SeqVar{\Object}'}
  \).
  (Context-hole filling in a well formed context can only
  descend inside an arrangement that is alone in its equivalence
  class.
  This is part of the motivation for our requirement that
  $\ArrangementEquiv$ must not relate
  distinct arrangements containing holes.)
\item
  \(
      \FillContextAux{\Object}{\SeqVar{\Object}}
    = \Pair{\Object}{\SeqVar{\Object}}
  \)
  if $\Object$ is not a context.
  (This is the only way context-hole filling can skip over embedded
  objects which are non-singleton equivalence classes of
  arrangements.)
\item
  $\FillContextAux{\Symbol}{\SeqVar{\Object}}=\Pair{\Symbol}{\SeqVar{\Object}}$
  and
  $\FillContextAux{n}{\SeqVar{\Object}}=\Pair{n}{\SeqVar{\Object}}$.
\item
  Context-hole filling essentially traverses the arrangement tree in
  a left-to-right order filling in holes in the order it encounters
  them. 
  Thus, for any arrangements $\Arrangement$, $\Arrangement_1$, and
  $\Arrangement_2$, core arrangement $\CoreArrangement$, and object
  sequences $\SeqVar{\Object}_1$, $\SeqVar{\Object}_2$,
  $\SeqVar{\Object}_3$, and $\SeqVar{\Object}_4$, if it holds that

  $$
    \begin{array}{@{}l@{\;=\;}l@{\qquad}l@{\;=\;}l@{}}
          \FillContextAux{\Arrangement}{\SeqVar{\Object}_1}
        & \Pair{\Arrangement'}{\SeqVar{\Object}_2}
      &
          \FillContextAux{\Arrangement_1}{\SeqVar{\Object}_2}
        & \Pair{\Arrangement_1'}{\SeqVar{\Object}_3}
    \\
          \FillContextAux{\Arrangement_2}{\SeqVar{\Object}_3}
        & \Pair{\Arrangement_2'}{\SeqVar{\Object}_4}
      &
          \FillContextAux{\CoreArrangement}{\SeqVar{\Object}_2}
        & \Pair{\CoreArrangement'}{\SeqVar{\Object}_3}
    \end{array}
  $$
  then all of these must follow:
  $$
    \begin{array}{@{}l@{\;=\;}l@{\qquad}l@{\;=\;}l@{}}
          \FillContextAux{\Arrangement\CoreArrangement}{\SeqVar{\Object}_1}
        & \Pair{\Arrangement'\CoreArrangement'}{\SeqVar{\Object}_3}
        &
         \FillContextAux{\Arrangement^{\Arrangement_1}_{\Arrangement_2}}{\SeqVar{\Object}_1}
        & \Pair{\Arrangement'{}^{\Arrangement_1'}_{\Arrangement_2'}}{\SeqVar{\Object}_4}
    \\
          \FillContextAux{\Arrangement^{\Arrangement_1}}{\SeqVar{\Object}_1}
        & \Pair{\Arrangement'{}^{\Arrangement_1'}}{\SeqVar{\Object}_3}
      &
          \FillContextAux{\Arrangement_{\Arrangement_1}}{\SeqVar{\Object}_1}
        & \Pair{\Arrangement'{}_{\Arrangement_1'}}{\SeqVar{\Object}_3}
    \\
          \FillContextAux{\overline{\Arrangement}}{\SeqVar{\Object}_1}
        & \Pair{\overline{\Arrangement'}}{\SeqVar{\Object}_2}
      &
         \FillContextAux{\underline{\Arrangement}}{\SeqVar{\Object}_1}
        & \Pair{\underline{\Arrangement'}}{\SeqVar{\Object}_2}
    \end{array}
  $$
  
 \item $\FillContextAux{P_\Object}{\SeqConcat{\SeqVar{\Object_1}}{\SeqVar{\Object_2}}}=\FillContextAux{P_{\Object_3}}{\SeqVar{\Object_2}}$ if $\FillContextAux{\Object}{\SeqConcat{\SeqVar{\Object_1}}{\SeqVar{\Object_2}}}=\FillContextAux{{\Object_3}}{\SeqVar{\Object_2}}$. (Context-hole filling descends object pointers until it encounters a hole)
\end{enumerate}

\note{}

\note{

}

\note{}

\begin{example}
  \label{ex:context-hole-filling}
  Here are some examples of context-hole filling
  :
  $$
    \setlength{\nextarrayminrowskip}{1.5ex}
    \begin{array}{@{}r@{\;=\;}l@{\qquad}@{\qquad}r@{\;=\;}l@{}}
       ({\ContextHole}\,{\ContextHole})\Seq{{\Object},{\Object}} & {\Object}\,{\Object}
    & \Mark{\ContextHole}\Seq{\Object}                            & \Mark{\Object}
   
    \\ \FillContext{\FunTypeP{\ContextHole}{\Object_1}}{\FunType{\Object_2}{\Object_2}}
       & \FunType{\FunTypeP{\Object_2}{\Object_2}}{\Object_1}
    &   (\Subst{\ContextHole}{\ContextHole}{\ContextHole})\Seq{{\Object_1},{\Object_2},{\Object_3}}
       & (\Subst{\Object_1}{\Object_2}{\Object_3})
    \end{array}
    \eqno
    \EEM
  $$
\end{example}

We now will define $\ContextSet{\Parens{\SetS_1,\SetS_2}}$ to be the the contexts which act as functions from $\SetS_1$ to $\SetS_2$, i.e., the set of every
context $\Object_{\mathsf{c}}$ of arity 1 such that for all
$\Object\in\SetS_1$ it holds that
$\Object_{\mathsf{c}}\Seq{\Object}\in\SetS_2$.
Let $\ContextSet{\SetS}=\ContextSet{\Parens{\SetS,\SetS}}$.

\note{}

Given a relation $\RelationR$ such that
$(\Domain{\RelationR}\cup\Range{\RelationR})\subseteq\SetS\subseteq\ObjectSet$,
let $\CompatibleClosure[\SetS]{\RelationR}$ denote the
\DefiningUse{$\SetS$-compatible closure of $\RelationR$}, defined as follows:
if $\Object_{\mathsf{c}}\in\ContextSet{\SetS}$ and
$\Object_1\GroundRelation{\RelationR}{}\Object_2$,\footnote{
$\Object_1\GroundRelation{\RelationR}{}\Object_2$ is an alternative notation for $\Pair{\Object_1}{\Object_2}\in\RelationR$. See appendix A.4 for details.}
 then
\(
  \FillContext{\Object_{\mathsf{c}}}{\Object_1}
  \CompatibleRelation[\SetS]{\RelationR}{}
  \FillContext{\Object_{\mathsf{c}}}{\Object_2}
\).
Let $\CompatibleClosure[]{\RelationR}$ denote
$\CompatibleClosure[\SetS]{\RelationR}$ for some set $\SetS$ which
the reader can infer from the context of discussion.

\note{}

Let $c$ range over \DefiningUse{primitive constructors},  non-hole objects whose only
immediate subobjects are $\ContextHole$.
\note{}

\begin{example}
  Here are some examples of primitive constructors \cite[p 134]{esopcallneed}, \cite[p386]{metalambda}, \cite[pg 360]{multistage}, \cite{8005074}:
  $$
    \begin{array}[b]{@{}l@{\qquad}l@{\qquad}l@{\qquad}l@{\qquad}l@{\qquad}l@{\qquad}l@{}}
      {({\ContextHole}\,{\ContextHole})}
    &
      {^{\ContextHole}\downarrow\ContextHole\cdot \ContextHole}
    &
      {!{\ContextHole}}
      &
      \langle\ContextHole\rangle
    &
      {{\ContextHole}+{\ContextHole}}
    &
      {{\ContextHole}={\ContextHole}\in{\ContextHole}}
    \end{array}
    \eqno
    \EEM
  $$
\end{example}

Every well formed non-hole object $\Object$ can be decomposed into a
primitive constructor and the subobjects to be placed in the primitive
constructor's holes.\note{}
A \DefiningUse{primitive constructor decomposition} of $\Object$ is a
pair $\Pair{\Constructor}{\SeqVar{\Object}}$ such that
$\Object=\Constructor\SeqVar{\Object}$.
An object will have one primitive constructor decomposition for each
of its arrangements.
Furthermore, the subobjects in a decomposition can be recursively
decomposed similarly.
A recursive decomposition of an object into primitive constructors is
very similar to the concept of an \FirstUse{abstract syntax tree} of a
string in a language defined by a grammar.
If any of the equivalence classes in an object are non-singletons,
then the object will not have a unique recursive decomposition.

\begin{example}
  Some examples of recursive decomposition of an object into
  primitive constructors can already be seen in
  \autoref{ex:context-hole-filling}.
  Here are some additional examples:
  $$\langle\Parens{!{\Object}}\rangle
      = \FillContext{\langle{\ContextHole}\rangle}
                    {\FillContext{!{\ContextHole}}
                                 {\Object}}
    \qquad
        {({\Object_1}+{\Object_2})}+{\Object_3}
      =\FillContext{({\ContextHole}+{\ContextHole})}
                    { \FillContext{({\ContextHole+}{\ContextHole})}
                                  {{\Object_1},{\Object_2}}
                     ,\Object_3}$$
\end{example}

\note{}

\subsection{Syntax Shorthand: Primitive Constuctor Decomposition}

\Autoref{conv:arrangement-object-coercion} allows avoiding the
need to write $\ArrangementEquivClass{\:\cdot\:}$ by 
implicitly invoking $\ArrangementEquivClass{\:\cdot\:}$ at obvious
arrangement boundaries and also at most uses of $\Parens{\:\cdot\:}$ in
arrangements.
For example, \autoref{conv:arrangement-object-coercion} allows us to
know that the expression
$\App{\AppP{\Object_1}{\Object_2}}{\Object_3}$ stands for the object
\note{}
whose primitive constructor decomposition is given by
$\Object'=\Constructor_{\AppSym}\Seq{\Constructor_{\AppSym}\Seq{{\Object_1},{\Object_2}},\Object_3}$
where $\Constructor_{\AppSym}=\App{\HoleW}{\HoleW}$.
\note{}

But the shorthand notation provided by
\autoref{conv:arrangement-object-coercion} is not enough.
Additionally, we want to allow inferring uses of
$\ArrangementEquivClass{\:\cdot\:}$ in other places in the middle of
what appear to be arrangements.
As a concrete example, we want to allow inferring that the expression
$\App{\App{\Object_1}{\Object_2}}{\Object_3}$ stands for the same
object as the expression
$\App{\AppP{\Object_1}{\Object_2}}{\Object_3}$, namely the object
$\Object'$ mentioned in the previous paragraph.
We want that the expression
$\App{\App{\Object_1}{\Object_2}}{\Object_3}$ must \emph{not} stand
for the object whose primitive constructor decomposition is
$\Constructor''\Seq{{\Object_1},{\Object_2},{\Object_3}}$ where
$\Constructor''=({\HoleW}\mathbin{\AppSym}{\HoleW}\mathbin{\AppSym}{\HoleW})$.

To provide the additional shorthand notation that is needed, we
establish mechanisms for (1)~declaring primitive constructors and
(2)~parsing arrangements.
We build the parsing mechanism by adapting the notions of operator
precedence
and declared associativity from parsing of languages to our setting;
this will allow splitting what appears to be a single primitive
constructor into multiple primitive constructors.

As an auxiliary device, we define splicing of arrangements.%
\note{}
Remember that every arrangement is, in effect, a sequence of core
arrangements (symbols, objects, numbers, or \note{}overlined
arrangements), possibly superscripted or subscripted.%
\note{}
An arrangement $\Arrangement'$ can be \DefiningUse{spliced} into
another arrangement $\Arrangement''$ by inserting the main core
arrangement sequence of $\Arrangement'$ into one of the core
arrangement sequences of $\Arrangement''$ in place of an occurrence of
$\ContextHole$.

\begin{convention}[Declaring and Parsing Primitive Constructors]
  \label{conv:declaring-parsing-constructors}
  \begin{enumerate}
  \item
    Unless prevented by part
    \ref{conv:declaring-parsing-constructors:parsing} of this
    convention, at the first use of a proper arrangement
    $\Arrangement$, if there is a primitive constructor
    $\Constructor=\Set{\Arrangement'}$ and objects $\Object_1$,
    $\ldots\,$, $\Object_n$ such that
    $\Set{\Arrangement}=\FillContext{\Constructor}{\Object_1,\ldots,\Object_n}$,
    then this use of $\Arrangement$ \DefiningUse{declares} the
    primitive constructor $\Constructor$ and the arrangement
    $\Arrangement'$.
    Note that $\Arrangement'$ differs from $\Arrangement$ exactly in
    having $\ContextHole$ in place of every non-$\ContextHole$ object
    appearing in $\Arrangement$.
  \item
    \label{conv:declaring-parsing-constructors:parsing}
    At each place where we coerce an arrangement $\Arrangement$ into
    an object $\Object$ using
    \autoref{conv:arrangement-object-coercion}, the arrangement
    $\Arrangement$ is inspected to see if it can be built by splicing
    together already-declared arrangements.
    If $\Arrangement$ can be built entirely by splicing together
    already-declared arrangements, and then filling the holes in the
    splicing result with objects, and there is no explicit indication
    forbidding the use of this convention, then $\Arrangement$ is to
    be interpreted as though it had been written with uses of
    $\ArrangementEquivClass{\:\cdot\:}$ around each splice point.
    If there is more than one way $\Arrangement$ can be built by
    splicing already-declared arrangements, then it must be specified
    somewhere which one to choose.
    (This choice will typically involve notions of operator precedence
    and declarations of associativity.)
  \end{enumerate}
\end{convention}

\begin{example}
  Suppose we have written the expressions $\langle\Object_1\rangle$
  and $!{\Object_2}$.
  This declares the primitive constructors
  $\langle\Object_1\rangle$ and $!{\Object_2}$.
  If we then write $\Object=\langle!{\Object'}\rangle$, then by
  \autoref{conv:declaring-parsing-constructors} this produces the same
  result as writing
  $\Object=\langle{\Parens{!{\Object'}}}\rangle$.
  This happens because the arrangement
  $\langle!{\Object'}\rangle$ can be built by splicing
  $!{\ContextHole}$ into $\langle\Object_1\rangle$ and then
  filling the hole with~$\Object'$.

  (If we wanted to avoid the interpretation of
  \autoref{conv:declaring-parsing-constructors}, we could do so by
  avoiding the implicit coercion of
  \autoref{conv:arrangement-object-coercion} and writing instead
  $\Object=\ArrangementEquivClass{\langle{!{\Object'}}\rangle}$,
  which would use the primitive constructor
  $\langle{!{\ContextHole}}\rangle$ instead of the two smaller
  primitive constructors $\langle{\ContextHole}\rangle$ and
  $!{\ContextHole}$.)

  Suppose we write the expression $\App{\Object_1}{\Object_2}$.
  This declares the primitive constructor
  $\Constructor_{\AppSym}=\App{\HoleW}{\HoleW}$.
  If we then state that $\Constructor_{\AppSym}$ is left-associative,
  then writing $\Object=\App{\App{\Object_1}{\Object_2}}{\Object_3}$
  produces the same result as writing
  $\Object=\App{\AppP{\Object_1}{\Object_2}}{\Object_3}$.
  If we did not give the associativity of $\Constructor_{\AppSym}$,
  then writing $\Object=\App{\App{\Object_1}{\Object_2}}{\Object_3}$
  would be an error, because there are multiple distinct ways the
  arrangement $\App{\App{\ContextHole}{\ContextHole}}{\ContextHole}$
  can be built by splicing the arrangement
  $\App{\ContextHole}{\ContextHole}$ into itself.
\end{example}

\note{}

\subsection{Names, Binding, $\upalpha$-Conversion, and Substitution}
\label{sec:names-binding-alpha-conversion-substitution}

\note{}
The relation $\ArrangementEquiv$ provides a mechanism for working with
syntax considered modulo equivalences on arrangements.
One of the most important equivalences is the notion of
\FirstUse{$\upalpha$-conversion} which renames \FirstUse{bound names}.\footnote{
We do not give an especially sophisticated notion of binding here.
We are only interested in providing a concept of binding that can be readily grasped and is sufficiently general for wide use in a variety of grammars.
The notion of equivalence we provide is intended to be used in defining other syntactic equivalences in addition to $\alpha$-equivalence.
}
\note{}

Some of the members of $\ObjectSet$ can be declared to be
\FirstUse{names}.
The names may be furthermore subdivided into groups.
Formally, the concepts of names and groups of names are given by an
equivalence relation
${\NamesInSameGroupRel}\subset\ObjectSet\times\ObjectSet$ which relates
names in the same group.
An object $\Object$ is a \DefiningUse{name} iff
$\Object\NamesInSameGroupRel\Object$.
Declaring a subset $\SetS\subset\Object$ to be a \DefiningUse{name
  group} is the same as declaring that $\SetS$ is a
$\NamesInSameGroupRel$-equivalence class\note{}.
The definition of $\NamesInSameGroupRel$ will be extended
incrementally with declarations of groups.
Any objects that have not been declared to be related by
$\NamesInSameGroupRel$ are \emph{not} related by
$\NamesInSameGroupRel$.
To keep things simple we require that no name contains another name
(of the same group or of a different group) as a subobject.

Specific primitive constructors can be declared to
\DefiningUse{bind} a name placed in one of the constructor's holes across some of the constructor's holes.
We define the \DefiningUse{free names} of an object $\Object$, written
$\FreeVars{\Object}$:
\begin{enumerate}
\item
  If $\Object$ is a name, then $\FreeVars{\Object}=\Set{\Object}$.
\item
  Otherwise, if $\FreeVars{\Object}$ is defined, it is as follows.

  First, we must define the free names of primitive constructor
  decompositions (p.c.d.'s) of $\Object$.
  Suppose
  $\Object=\FillContext{\Constructor}{\Object_1,\ldots,\Object_n}$
  gives one such p.c.d.\
  Let $\SetS_i$ be the names bound by $\Constructor$ in $\Object_i$
  for $1\leq{i}\leq{n}$.
  Then
  $\FreeVars{c,\Seq{\Object_1,\ldots,\Object_n}}=\bigcup_{i\in\Set{1,\ldots,n}}\FreeVars{\Object_i}\setminus\SetS_i$.

  If there exists a set $\SetS$ such that
  $\SetS=\FreeVars{\Constructor,\SeqVar{\Object}}$ for every p.c.d.\ $\Pair{\Constructor}{\SeqVar{\Object}}$ of $\Object$,
  then $\FreeVars{\Object}=\SetS$.
  
\end{enumerate}
The free names of an arrangement $\Arrangement$ are defined by
$\FreeVars{\Arrangement}=\FreeVars{\Set{\Arrangement}}$.
A name that is not free is \DefiningUse{bound}.

\note{}

\begin{example}
  Consider
  $\Constructor_{\upsflambda}=\Abs{\ContextHole}{\ContextHole}$ of
  arity 2.
  Suppose we declare that $\Constructor_{\upsflambda}$ binds any name
  placed in its first hole in both of its holes.
  Suppose we declare that
  $\AllSuchThat{\ConcreteTermVar{i}}{i\in\NatNumSet}$ is a name group.
  (We will in fact make both of these declarations later, so this
  example is not just hypothetical.)
  Suppose that we have not declared any bindings for the constructor
  $\Constructor_{\AppSym}=\App{\HoleW}{\HoleW}$.
  Then
  \(
      \FreeVars{\App{\AbsP{\ConcreteTermVar{1}}
                          {\AppP{\ConcreteTermVar{1}}
                                {\ConcreteTermVar{2}}}}
                    {\ConcreteTermVar{3}}}
    = \Set{{\ConcreteTermVar{2}},{\ConcreteTermVar{3}}}
  \).

  Consider
  $\Constructor_{\mathsf{let}}=\LetInP{\ContextHole}{\ContextHole}{\ContextHole}$
  of arity 3.
  Suppose we declare that $\Constructor_{\mathsf{let}}$ binds any name placed
  in its 1st hole in its 1st and 3rd hole.
  Then
  \(
      \FreeVars{\LetIn{\ConcreteTermVar{1}}
                      {\ConcreteTermVar{3}}
                      {\AppP{\ConcreteTermVar{1}}
                            {\ConcreteTermVar{2}}}}
    = \Set{{\ConcreteTermVar{2}},{\ConcreteTermVar{3}}}
  \).
\end{example}

\note{}

\note{}

We now define the auxiliary notion of \DefiningUse{name swapping}.
Given two names $\Object_{\mathsf{x}}$ and $\Object_{\mathsf{y}}$ such
that $\Object_{\mathsf{x}}\NamesInSameGroupRel\Object_{\mathsf{y}}$,
let
$\SwapNamesIn{\Object_{\mathsf{x}}}{\Object_{\mathsf{y}}}{\Object}$ be
the object $\Object'$ that results from replacing every occurrence of
$\Object_{\mathsf{x}}$ in $\Object$ by $\Object_{\mathsf{y}}$, and
\emph{vice versa}.
Let
$\SwapNamesIn{\Object_{\mathsf{x}}}{\Object_{\mathsf{y}}}{\Arrangement}$
be defined similarly.

\note{}

We now define \DefiningUse{$\upalpha$-conversion}.
Let $\AlphaConvertibleRel$ be the smallest equivalence relation
satisfying the following condition.
For all $\Object_{\mathsf{x}}$, $\Object_{\mathsf{y}}$, $\Object$, and
$\Arrangement$, if
$\Object_{\mathsf{x}}\NamesInSameGroupRel\Object_{\mathsf{y}}$ and
\(
    \Set{\Object_{\mathsf{x}},\Object_{\mathsf{y}}}\cap\FreeVars{\Object}
  = \Set{\Object_{\mathsf{x}},\Object_{\mathsf{y}}}\cap\FreeVars{\Arrangement}
  = \emptyset
  \),
\relax
then
$\Object\AlphaConvertibleRel\SwapNamesIn{\Object_{\mathsf{x}}}{\Object_{\mathsf{y}}}{\Object}$
and
$\Arrangement\AlphaConvertibleRel\SwapNamesIn{\Object_{\mathsf{x}}}{\Object_{\mathsf{y}}}{\Arrangement}$.

\note{}

\begin{definition}[$\upalpha$-Conversion as a Syntactic Equivalence]
  \label{def:alpha-convertible-implies-syntactically-equivalent}
  If a paper says that it is ``working modulo $\alpha$" or ``identifying $\alpha$-equivalent terms" that means
   $\AlphaConvertibleRel$ restricted to arrangements is
  a subset of $\ArrangementEquiv$, i.e., if
  $\Arrangement_1\AlphaConvertibleRel\Arrangement_2$ then
  $\Arrangement_1\ArrangementEquiv\Arrangement_2$.
\end{definition}

\note{}

\Autoref{def:alpha-convertible-implies-syntactically-equivalent}
implies that $\ArrangementEquiv$ will change whenever adjustments are
made to the declared bindings of primitive constructors or to the
definition of $\NamesInSameGroupRel$.

We now define the \DefiningUse{substitution} operation, written as
$\MetaSubst{\Object}{\Object_{\mathsf{x}}}{\Object'}$.
This expression will be defined to stand for the result of replacing
all free occurrences of $\Object_{\mathsf{x}}$ in $\Object$ by
$\Object'$.
This operation must be defined carefully.
The result of $\MetaSubst{\Object}{\Object_{\mathsf{x}}}{\Object'}$
must not allow names that are free in $\Object'$ to be captured by
bindings in $\Object$.
Also, the operation must respect $\ArrangementEquiv$ so that if both
$\Object$ and $\Object'$ are well formed, then
$\MetaSubst{\Object}{\Object_{\mathsf{x}}}{\Object'}$ is also well
formed.
Given a name $\Object_{\mathsf{x}}$, define
$\MetaSubst{\Object}{\Object_{\mathsf{x}}}{\Object'}$ formally as
follows.
\begin{enumerate}
\item
  If ${\Object}={\Object_{\mathsf{x}}}$, then
  $\MetaSubst{\Object}{\Object_{\mathsf{x}}}{\Object'}={\Object'}$.
\item
  Otherwise, $\MetaSubst{\Object}{\Object_{\mathsf{x}}}{\Object'}$ is defined as
  follows.

  First, we must define substitution for primitive constructor
  decompositions (p.c.d.'s).
  Given $\Object=\Constructor\Seq{\Object_1,\ldots,\Object_n}$, let
  $\SetS$ be the subset of $\Set{\Object_1,\ldots,\Object_n}$ of names
  bound by this occurrence of $\Constructor$.
  If $\SetS\cap\FreeVars{\Object'}\neq\emptyset$, then let
  $\MetaSubst{\Pair{\Constructor}{\Seq{\Object_1,\ldots,\Object_n}}}{\Object_{\mathsf{x}}}{\Object'}$
  be undefined.\footnote{For simplicity, we do not check whether the
    substitution needs only to proceed into holes of $\Constructor$
    which are not subject to its bindings.
    This will behave well enough for our uses provided each group of
    names is big enough that fresh names can  be found.}
  Otherwise, let
  \(
      \MetaSubst{\Pair{\Constructor}{\Seq{\Object_1,\ldots,\Object_n}}}{\Object_{\mathsf{x}}}{\Object'}
    =
      \FillContext
        {\Constructor}
        {\MetaSubst{\Object_1}{\Object_{\mathsf{x}}}{\Object'},
         \ldots,
         \MetaSubst{\Object_n}{\Object_{\mathsf{x}}}{\Object'}}
  \).
  {\sloppy\par}%

  If there exists an $\Object''$ such that
  $\Object''=\MetaSubst{\Pair{\Constructor}{\SeqVar\Object}}{\Object_{\mathsf{x}}}{\Object'}$
  for every p.c.d.\
  $\Pair{\Constructor}{\SeqVar{\Object}}$ of $\Object$ such that
  $\MetaSubst{\Pair{\Constructor}{\SeqVar\Object}}{\Object_{\mathsf{x}}}{\Object'}$
  is defined, then
  $\MetaSubst{\Object}{\Object_{\mathsf{x}}}{\Object'}=\Object''$.
  Otherwise $\MetaSubst{\Object}{\Object_{\mathsf{x}}}{\Object'}$ is
  undefined.\footnote{So the substitution must be defined for at least
    one of the primitive constructor decompositions to get a defined
    result.}
\end{enumerate}

\note{}

\note{}

\note{}

\note{}

\note{}

\begin{example}
  \label{ex:objects}
  Below, on the left are some example syntactic objects \cite[p 134]{esopcallneed}, \cite{8005074}, \cite[p 386]{metalambda}.
    These objects may not be well formed, because the singleton
  sets may not be equivalence classes of $\ArrangementEquiv$.
  The objects to the right of them are adjusted to be well formed (assuming the
  subobjects $\Object_1$, to $\Object_4$ are
  well formed):
  $$
    \begin{array}{lcl}
       \Set{\lambda P_{\Object_1}.P_{\Object_2}}&&\ArrangementEquivClass{\lambda P_{\Object_1}.P_{\Object_2}}
\\ \Set{\Pi{P_{\Object_1}}:{P_{\Object_2}}.{P_{\Object_3}}}
  &&\ArrangementEquivClass{\Pi{P_{\Object_1}}:{P_{\Object_2}}.{P_{\Object_3}}}    
    \\ \Set{{}^{P_{\Object_1}}\downarrow{\Set{P_{\Object_2}\, P_{\Object_3}}}\cdot P_{\Object_4}}
    &&\ArrangementEquivClass{{}^{P_{\Object_1}}\downarrow
    \ArrangementEquivClass{P_{\Object_2}\, P_{\Object_3}}\cdot P_{\Object_4}}
    \end{array}
  $$%
\end{example}

\subsection{Production Rules for Defining Syntactic Sets}
\label{sec:production-rules}

We have already defined syntactic objects, but the set $\ObjectSet$ is
too big.
Carefully defined subsets of $\ObjectSet$ may be defined via \DefiningUse{syntax
production rules}, which we write in the form
$$
  \MetaVar_1,\ldots,\MetaVar_n
  \in \SetS
  \mathrel{\GrammarSym}
         \Alternative_1
    \mid \cdots
    \mid \Alternative_m
$$
where
$\MetaVar_1$, $\ldots\,$, $\MetaVar_n$ are metavariables,
$\SetS$ is the name of the subset of $\ObjectSet$ being defined, and
$\Alternative_1$, $\ldots\,$, $\Alternative_m$ are \DefiningUse{alternatives}.
Each alternative is either the special notation ``$\cdots$'' or an
expression $e$, together with an optional side condition $c$ (written ``$e\quad\mathrm{if\,}c$", where $c$ is a formula containing only expressions), which evaluates to a member of $\ObjectSet$ when values are
supplied for metavariables occurring in $e$, provided both $c$  holds of that choice of metavariables.
One can omit the ``$\in \SetS$'', allowing the reader to fill in $\SetS$ whose name is distinct
from the names of all other declared sets.
One can omit the side condition in which case we can read it as $\mathrm{if\, true}$.
One can provide a global side condition $\mathrm{if\,}c'$ which we read as appending $\wedge c'$ to all $\Alternative$. 

\note{}

Such a syntax production rule has the following effects:
\begin{enumerate}
\item
  \note{}
  It declares $\SetS$ to be a set of syntactic objects, in particular
  the smallest one that
  satisfies all other constraints placed on it not just by this rule
  but also by the rest of the document.
\item
  It declares the metavariables $\MetaVar_1$, $\ldots\,$, $\MetaVar_n$
  to range over the set $\SetS$.
  \item A global side condition $\mathrm{if\,}c'$ appends $\wedge c'$ to each $\Alternative_1,\ldots\Alternative_n$.
  \item 
  If each $\Alternative_1,\ldots\Alternative_n$ contains only undecorated instances of $\MetaVar$, then for any  $\Alternative$ containing multiple instances of $\MetaVar$ and no side conditions containing $\MetaVar$ that apply to $\Alternative$, we can rewrite it with each $\MetaVar$ given a different decoration.
  I.e., $m\in M\mathrel{::=}x\mid m\, m$ becomes $m, m_1,m_2\in M\mathrel{::=}x\mid m_1\, m_2$.

\item
  For each alternative $\Alternative$ in the rule which is not
  ``$\cdots$'', a constraint on the membership of $\SetS$ is added.
  The constraint is that for each legal choice\footnote{
  By legal choice we mean a choice of metavariables matching the sets they are declared to range over and fulfilling any constraints added by any side conditions.}
   of values for the
  metavariables occurring in $\Alternative$,
  if $\Object$ is the result of evaluating the expression $e$ in $\Alternative$ using those
  metavariable assignments, then $\Object\in\SetS$.

  Metavariables occurring in an alternative $\Alternative$ that are
  not yet declared to range over any set are presumed 
  to range over a countable set of objects disjoint from all the other sets of objects in the paper.
  This assumption is dropped if a value for a metavariable gets declared later in the paper and values for $\Alternative$ are recalculated accordingly.
\item
  If the first alternative is not the special alternative
  ``$\cdots$'', then any constraints on the membership of $\SetS$
  established by earlier rules are forgotten.
\item
  The rule triggers a recalculation of \emph{all} of the sets declared
  by all syntax production rules.
  Such a recalculation is also triggered whenever the definition of
  $\ArrangementEquiv$ is altered.
  Or when a definition of what metavariables range over is altered.

  This recalculation evaluates all of the constraint expressions for
  all syntactic sets using the \emph{current} bindings for all
  metavariables, set names, the equivalence relation
  $\ArrangementEquiv$, etc., and rebinds the set names to the
  recalculated values in the subsequent text.\footnote{%
    It is an error if there is not a unique assignment of smallest
    values to the declared sets.
    Normally, the existence of a unique assignment will be provable
    using a fixed point theorem like the Knaster-Tarski theorem.
    However, the notation allows putting strange side conditions in
    the constraint expressions in alternatives, and this can cause a
    failure.
    }

\end{enumerate}

Multiple rules can be given for the same set $\SetS$.
If a later rule for $\SetS$ begins with the special alternative
``$\cdots$'', then its alternatives are combined with the alternatives
already in force for $\SetS$.
 Usually the alternatives of the later
rule replace the previous alternatives if this is not the case.
However, if the author uses a single alternative in each of their production rules, then they normally expect these to be combined as though they had used $\cdots$.
The special alternative ``$\cdots$'' used as the final alternative of
a rule has no mathematical consequence
and is used only as a signal to the reader warning that there will be
later rules for the same set.

\note{}

\note{}

\note{
}

When a syntax alternative is intended to allow building terms from
multiple subterms of the same set, it is necessary to use distinct
metavariables for each possible subterm to allow the subterms to
differ.
It is always possible to find distinct metavariables for the same set
by using subscripts.

\begin{example}
  We can define the usual \DefiningUse{simple types} like this:
  $$
    \begin{array}{r@{\;\in\;}l@{\;\mathrel{\GrammarSym}\;}l}
       \TyVarA, \TyVarB & \TyVarSet      & \ConcreteTyVar{i}
    \\ \Type            & \SimpleTypeSet & \TyVarA \mid \FunType{\Type_1}{\Type_2}
    \end{array}
  $$
  Given this definition, a possible example type is
  $\Type_0=\FunType{\TyVarA}{\FunTypeP{\TyVarB}{\TyVarA}}$.
  In this example $\Type_0$, we leave unspecified which exact type
  variables are used.
  We could make $\Type_0$ concrete by specifying
  $\TyVarA=\ConcreteTyVar{0}$ and $\TyVarB=\ConcreteTyVar{1}$ yielding
  $\Type_0=\FunType{\ConcreteTyVar{0}}{\FunTypeP{\ConcreteTyVar{1}}{\ConcreteTyVar{0}}}$.
  If we had written the second alternative in the production rule for
  $\SimpleTypeSet$ as $\FunType{\Type}{\Type}$, then the type $\Type_0$
  would not be allowed and we could only write types like
  $\FunType{\TyVarA}{\TyVarA}$ and
  $\FunType{\FunTypeP{\TyVarA}{\TyVarA}}{\FunTypeP{\TyVarA}{\TyVarA}}$
  where both arguments of each $\FunType{}{}$ are equal.
  \note{}
\end{example}

\note{}

\note{}
\begin{example}We can define the \emph{lambda calculus} like this:
$$
e\in \mtnormsf{exp}\mathrel{::=}v\mid \lambda v.e \mid e\, e
$$
Each $v$ ranges over a countable set of objects disjoint from the objects produced by the other production rules.
The production rule $e\in \mtnormsf{exp}\mathrel{::=}v$ can be read as giving us the constraint $\mtnormsf{var}\subseteq \mtnormsf{exp}$.
 The constraint  $\AllSuchThat{\ArrangementEquivClass{\lambda P_v.P_e}}{\mtnormsf{ptr}(v)=P_v\wedge\mtnormsf{ptr}(e)=P_e}\subseteq\mtnormsf{exp}$ is given by $e\in \mtnormsf{exp}\mathrel{::=}\lambda x.e$.
 The constraint  $\AllSuchThat{\ArrangementEquivClass{ P_{e_1}\, P_{e_2}}}{\mtnormsf{ptr}(e_1)=P_{e_1}\wedge \mtnormsf{ptr}(e_2)=P_{e_2}}\subseteq\mtnormsf{exp}$ is given by $e\in \mtnormsf{exp}\mathrel{::=}e\, e$.
We pick the least $\mtnormsf{exp}\subseteq \ObjectSet$ and $\mtnormsf{var}\subseteq\ObjectSet$  satisfying these constraints with an ordering given by the subset relation.

In addition to declaring $e$ as ranging over $\mtnormsf{exp}$ this definition also declares $e_1$, $e_2$,..., $e'$, $e''$ etc. to range over $\mtnormsf{exp}$ and similarly for $v\in\mtnormsf{var}$.
The subset of $\ObjectSet$ picked out by these constraints depends on the choice of equivalence relation $\ArrangementEquiv$, in the lambda calculus this is most likely $\alpha$ equivalence, although it may also be the identity relation on $\ArrangementSet$

In order to be confident that this set can be picked out (e.g. for $\mtnormsf{exp}$) we begin with $\mtnormsf{exp}^0=\emptyset$ and let $\mtnormsf{exp}^1$ contain all the things $\mtnormsf{exp}$  must contain if $\mtnormsf{exp}$ is at least $\mtnormsf{exp}^0$ and so on for each +1 case.
For a limit point $\varepsilon$ we let  $\mtnormsf{exp}^\varepsilon$ be $\bigcup\limits_{i=0}^\varepsilon\mtnormsf{exp}^i$. 
We take the least fixed point of the function $f\in\FunSpace{\mathcal{P}(\ObjectSet)}{\mathcal{P}(\ObjectSet)}$ such that $f(\mtnormsf{exp}^i)=\mtnormsf{exp}^{i+1}$ over some appropriately large set of $\mtnormsf{exp}^{i}$ ordered but the subset relation (this is smaller than $\mathcal{P}(\ObjectSet)$).

We define the reduction relation as the $e$-compatible closure of the smallest $\beta$ (in the ordering given by $\subseteq$) satisfying the constraint $(\lambda v.e_1)e_2\GroundRelation{\beta} {}(\MetaSubst{e_1}{v}{e_2})$.
The notation
$\MetaSubst{\Object_1}{\Object_2}{\Object_3}$ is defined in section 2.5.
We do the same for $\lambda v.e_1 v\GroundRelation{\eta}{} e$.
Note that because we have bracketed the term after substitution we are able to reapply equivalences that may have otherwise been lost in the process.
\end{example}

\begin{example}
Given the definition of simple types in Example 2.14 we can define the \emph{simply typed lambda calculus} as follows:
$$\hat{e}\in \mtnormsf{texp}\mathrel{::=}v\mid\hat{e}\,\hat{e}\mid\lambda v:\Type .\,\hat{e}$$ 
\note{}
 \end{example}

\note{}
\begin{example}
We can extend example 2.14 with records in a similar way to Pierce \cite[pg 129]{Pierce:2002:TPL:509043}. 
We can define \emph{lambda calculus with records}  like this:
$$
\begin{array}{rcl@{\qquad}rcl}
l\in\mtnormsf{label}&::=&\mtnormsf{y}_i &
R\in\mtnormsf{Type-Records}&::=&\epsilon\mid l:\Type, R\qquad\mathrm{where\,}l\notin\mtnormsf{lab}(R)
\\
\hat{e}\in\mtnormsf{texp}&::=&\cdots\mid\{r\}\mid\hat{e}.l
\\
t\in\mtnormsf{Record-Type}&::=&\Type\mid\{ R\}&r\in\mtnormsf{Term-Records}&::=&\epsilon\mid l=\hat{e}, r\qquad\mathrm{where\,}l\notin\mtnormsf{lab}(r)
\end{array}$$
Where we define $\mtnormsf{lab}$ s.t. $\mtnormsf{lab}(\epsilon)=\emptyset$, $\mtnormsf{lab}(l:\Type,R)=\{ l\}\cup\mtnormsf{lab}(R)$ and $\mtnormsf{lab}(l=\hat{e},r)=\{ l\}\cup\mtnormsf{lab}(r)$.
Both $r$ and $R$ are equivalent up to reordering (i.e. $l:\Type,R \ArrangementEquiv R,l:\Type $ and $l=\hat{e},r
\ArrangementEquiv r,l=\hat{e}$).
Here, $\ArrangementEquiv$ is the smallest equivalence relation fulfilling these constraints.
It is defined incrementally over each $R$ and each $r$ as a new one is added.

\note{
}
We add a rewriting rule:
$$
\{ l=v,r\}.l\GroundRelation{\mtnormsf{RCD}} {}v$$
For each $*\in \AllSuchThat{x\in\ObjectSet\times\ObjectSet}{x=\beta
}\cup \AllSuchThat{x\in\ObjectSet\times\ObjectSet}{x=\eta
}\cup\{\mtnormsf{RCD}\}$ we add additional constraints:
$$\begin{array}{c}
(\hat{e}_1\GroundRelation{*} {}\hat{e}_2)\\
\hline
(\hat{e}_1.l\GroundRelation{*} {}\hat{e}_2.l)
\end{array}
$$
$$\begin{array}{c}
(\hat{e}_1\GroundRelation{*} {}\hat{e}_2)\\
\hline
(\{ r,l=\hat{e}_1\}\GroundRelation{*} {}\{ r,l=\hat{e}_2\})
\end{array}
$$
(The horizontal line is read as the logical operator $\Rightarrow$).
\note{}
\end{example}

\note{}

\section{Model for Syntactic Math Text}

\note{}
\note{}
In this section we show that there is a model for SMT.
In order to do so, we choose sets to represent $\SymbolSet$, $\PositionSet$, $\PointerSet$, $\ContextHole$, $\epsilon$ and $\BarSign$.
Our invariant constraints are those that will hold of sets thought to approximate $\ObjectSet$ and $\ArrangementSet$ in the proof these are well defined.
Our constraints on the final selection will only hold of the set we pick out from these approximations.

\begin{definition}[Symbol, Pos, Pointer, $\{\ContextHole, \epsilon, \BarSign \}$] 
We can create a countable set, $\mtnormsf{D}$, representing symbols, accenting and positioning from the 
ordinals\footnote{With the Von Neumann encoding} following $\omega$ which are themselves smaller than $2\omega$.
We pick a finite set of elements, $\PositionSet$, from $\mtnormsf{D}$ to represent the positions subscript, superscript, pre-subscript, pre superscript, text above, text below etc (at least as many as positions as detailed in the  OpenDocument \cite{ISO26300} standard).
We pick out an element of $\mtnormsf{D}$ which we call $\BarSign$.
We pick out a element  $\mtnormsf{D}$ to represent the context-hole $\ContextHole$, and one to represent the empty arrangement $\epsilon$.
We let the remainder of the elements in $\mtnormsf{D}$ represent $\SymbolSet$ (at least as many symbols as in unicode).
\end{definition}

\begin{definition}[Invariant Constraints]

\noindent 

\noindent $\mtnormsf{ptr}\in\FunSpace{\ObjectSet}{\PointerSet}$ and $\mtnormsf{ptr}$ is a bijection between $\ObjectSet$ and $\PointerSet$.

\noindent $\BarSign\notin\ObjectSet\wedge \BarSign\notin\ArrangementSet $.

\noindent $\ContextHole\in\ObjectSet\wedge\ContextHole\notin\ArrangementSet\wedge\epsilon\in\ArrangementSet$

 \noindent $\PositionSet\mathrel{\bot}\ObjectSet\wedge \PositionSet\mathrel{\bot}\ArrangementSet $.

\noindent $ \SymbolSet\subset\CoreSet$.

 \noindent $ \NatSet\subset \CoreSet$.

\noindent $\CoreSet\subset\ArrangementSet$.
\end{definition}
 
\begin{definition}[Constraints on Final Selection]

\noindent

\noindent $ \PointerSet\subset\CoreSet$.

\noindent If $\Arrangement\in\ArrangementSet$, $\Arrangement\neq\epsilon$ and $x\in\SymbolSet$ then $\Tuple{\BarSign,x,\Arrangement}\in\CoreSet$.

\noindent $\ArrangementSet\times\CoreSet\subset\ArrangementSet$.

\noindent If
$\Arrangement \in\ArrangementSet$ and $x\in \FunSpace{\PositionSet} {\ArrangementSet\setminus\{\epsilon\}}$ and $x\neq\emptyset$ 
then

\noindent $\Pair{\Arrangement}{x}\in\ArrangementSet$.
 
\noindent If $\SetS\subset\ArrangementSet$ and $|\SetS|\leq\aleph_0$, then $ \SetS\in\ObjectSet
$.\footnote{We do not bother restricting objects to only include proper arrangements here as it does not particularly affect the logic of the proof.
Provided one can pick out unique members from $\SymbolSet$ for left parenthesis, right parenthesis and comma, its not too hard to express what it means for an arrangement to be proper with a logical formula.}
\end{definition}

\note{}
\begin{theorem}
$\ObjectSet$ and $\ArrangementSet$ are well defined.
\begin{proofsketch}
We define a sequence of sets thought to contain closer approximations of $\ObjectSet$ and $\ArrangementSet$ until some member contains a model for $\ObjectSet$ and $\ArrangementSet$ themselves.
The smallest set in our sequence contains all tuples of: \begin{enumerate}
\item  The set containing $\ContextHole$ (approximating $\ObjectSet$).
\item An injective function $p\in\FunSpace{\{\ContextHole\}}{\PointerSet}$ (approximating $\mtnormsf{ptr}$).
\item $\SymbolSet\cup\NatSet\cup\{\epsilon\}$ (approximating $\ArrangementSet$).
\item $\SymbolSet\cup\NatSet$ (approximating $\CoreSet$).
\end{enumerate}
Each subsequent set in our sequence contains those tuples of sets which would be added by applying our constraints as though $\ObjectSet$ were its approximation, $\PointerSet$ were its approximation and $\ArrangementSet$ were its approximation.
Where our sequence reaches a limit point each set in each tuple is calculated as though $\ArrangementSet$ was the union of arrangements up to that point (apart from the set approximating $\ArrangementSet$ which also gets the pointers to the approximation of $\ObjectSet$ at that limit).

These sets remain sufficiently small to pick mappings for $\ObjectSet$.
 Further, there is a fixed point for the function mapping each member of this sequence to the member above it.
 From this fixed point we can select a model for  $\ObjectSet$ and $\ArrangementSet$.
 
 (Full proof in Appendix B)
\end{proofsketch}
\end{theorem}

\section{How Can This Definition be Used?}
\note{
}
\paragraph{Non-MBNF ``Grammars"}
As well as covering some uses of MBNF to define syntax, SMT also provides us with a notion of what it means to use the structures of math-text together with syntactic equivalences, even in documents where MBNF does not feature, or where MBNF is mixed with other notation for picking out objects.
Coverage of this sort may require users to  select appropriate sets of objects that resolve ambiguities.

\paragraph{A Flexible Notion of Equivalence}
Not only is the notion of equivalence presented in SMT sufficient to deal with $\alpha$-equivalence over finite terms, regardless of how binding may be represented in the syntax, it also deals with things like equality up to reordering of finitely many chunks of syntax and equality of finitely many compositions with zero, both of which appear in the $\pi$-calculus \cite{Carbone}.
It deals with many of the equivalences an author might define using ``$=$,'' provided they do not quantify over an uncountable set when using it.
Furthermore it provides tools to consider equalities over sub-objects, not just the syntactic objects themselves, which can be vital when talking about the structure of a grammar.

\paragraph{Combining Objects in Math-Text}
SMT deals with most combinations of characters likely to appear in math text used to represent ``syntax" in a fairly general manner (it does not deal with matrices/grids, numbers other than the naturals, or an use of sets that cannot be thought of in terms of equivalences up to reordering and repetition on finite lists of elements, but none of these is likely to appear in ``syntax").
\note{}

\paragraph{Automatic Bracketing}
Since SMT preserves the tree-like structure of syntax, it can be readily used for grammars where authors treat bracketing as optional.
We also give authors the option of making this structure more explicit by primitive decomposition.
Bracketing structures may often also be derived by noticing where objects appear in production rules.

\paragraph{Functionality Inherited From BNF}
Our definition extends the basic functions covered by BNF to MBNF and the richer syntactic structures that are represented by math-text. 
Substitution of non-terminals becomes assigning values to metavariables and choice of production rules remains supported.

\paragraph{Hole Filling}
The following chunk of the MBNF we took from Chang and Felleisen \cite[p 134]{esopcallneed} defining $A$ can be handled by our definition using  \autoref{conv:arrangement-object-coercion}:$$
 \arraycolsep=2pt
 \begin{array}{rcl}
  e& =& \ConcreteTermVar{i}\mid \lambda \ConcreteTermVar{i} . e  \mid e \, e\\
  A& =&[\, ] \mid  A [ \lambda \ConcreteTermVar{i}. A ]\, e
 \end{array}
 $$

\note{}

\section{Related work}

OTT \cite{Sewell:2007:OET:1291220.1291155} provides a formal language for writing specifications like those written in MBNF.
The process of moving from an OTT specification to an MBNF can be performed automatically.
However, the focus of this article is moving the other way --- interpreting MBNF without requiring it to be specified in a theorem-prover friendly format.
Furthermore, we wish to provide a general mathematical intuition suitable for translation to multiple theorem provers, whereas OTT focuses on translating to COQ 8.3, HOL 4 and Isabelle directly, but offers less support for those seeking a general mathematical intuition.
In addition, OTT only supports context hole filling for contexts with a single hole and currently does not support rules being used coinductively. 
We already handle more cases of context hole filling and we aim to deal with coinduction, though SMT as it stands doesn't.

Guy Steele \cite{Steele:2017:TNO:3018743.3018773} covers many of the notational variants of BNF, including some MBNFs.
However, Steele's focus is primarily on surface differences.
 He does not discuss how the underlying mathematical structure of MBNF differs wildly from BNF.

Grewe et al.\ \cite{DBLP:journals/scp/GreweEPRM18} discuss the exploration of language specifications with first-order theorem provers.
However, they still require the reader to be able to intuitively translate language specifications to a sufficiently formal language first.
 This is the part of language specification checking this paper aims to help with.

Reynolds \cite[1-51]{Reynolds:2009:TPL:1611353} has the best attempt at a definition of MBNF, which he calls ``abstract syntax",\footnote{
We do not call MBNF ``abstract syntax," because some of it is concrete syntax.} 
which we could find after looking through the books in our collection.
However, he only deals context-free grammars and in many places he proceeds by example.

\section{Future Work}

\note{}

While we do not deal with trees of infinite breadth or depth here, we hypothesise that the method outlined in this document could be used on trees with countably infinite breadth and depth.
The main difference in doing so would be that $\ObjectSet$ and $\ArrangementSet$ would likely have  to be of cardinality $\aleph_2$, rather than $\aleph_1$, but apart from that it seems likely a similar proof would work.

\note{}

While we provide some powerful tools for writing syntax patterns more explicitly and dealing with numbers in the syntax, we do not provide procedures for generating countably many production rules.
Guy Steele \cite{Steele:2017:TNO:3018743.3018773} has done work in this area, but doesn't address differences between MBNF and BNF.
\note{}

\note{}

\bibliographystyle{eptcs}
 \bibliography{chapter,bibliography,conferences}

\begin{thebibliography}{10}
\providecommand{\bibitemdeclare}[2]{}
\providecommand{\surnamestart}{}
\providecommand{\surnameend}{}
\providecommand{\urlprefix}{Available at }
\providecommand{\url}[1]{\texttt{#1}}
\providecommand{\href}[2]{\texttt{#2}}
\providecommand{\urlalt}[2]{\href{#1}{#2}}
\providecommand{\doi}[1]{doi:\urlalt{http://dx.doi.org/#1}{#1}}
\providecommand{\bibinfo}[2]{#2}

\bibitemdeclare{inproceedings}{Carbone}
\bibitem{Carbone}
\bibinfo{author}{Marco \surnamestart Carbone\surnameend},
  \bibinfo{author}{Kohei \surnamestart Honda\surnameend} \&
  \bibinfo{author}{Nobuko \surnamestart Yoshida\surnameend}
  (\bibinfo{year}{2007}): \emph{\bibinfo{title}{Structured
  Communication-Centred Programming for Web Services}}.
\newblock In \bibinfo{editor}{Rocco \surnamestart De~Nicola\surnameend},
  editor: {\sl \bibinfo{booktitle}{Programming Languages and Systems}},
  \bibinfo{publisher}{Springer Berlin Heidelberg}, \bibinfo{address}{Berlin,
  Heidelberg}, pp. \bibinfo{pages}{2--17}.

\bibitemdeclare{inproceedings}{esopcallneed}
\bibitem{esopcallneed}
\bibinfo{author}{Stephen \surnamestart Chang\surnameend} \&
  \bibinfo{author}{Matthias \surnamestart Felleisen\surnameend}
  (\bibinfo{year}{2012}): \emph{\bibinfo{title}{The Call-by-need Lambda
  Calculus, Revisited}}.
\newblock In \bibinfo{editor}{Seidl}  \cite{esop2012}, pp.
  \bibinfo{pages}{128--147}.

\bibitemdeclare{inproceedings}{eberhart_et_al:LIPIcs:2015:5528}
\bibitem{eberhart_et_al:LIPIcs:2015:5528}
\bibinfo{author}{Clovis \surnamestart Eberhart\surnameend},
  \bibinfo{author}{Tom \surnamestart Hirschowitz\surnameend} \&
  \bibinfo{author}{Thomas \surnamestart Seiller\surnameend}
  (\bibinfo{year}{2015}): \emph{\bibinfo{title}{{An Intensionally
  Fully-abstract Sheaf Model for $\pi^\ast$}}}.
\newblock In \bibinfo{editor}{Lawrence~S. \surnamestart Moss\surnameend} \&
  \bibinfo{editor}{Pawel \surnamestart Sobocinski\surnameend}, editors: {\sl
  \bibinfo{booktitle}{6th Conference on Algebra and Coalgebra in Computer
  Science (CALCO 2015)}}, {\sl \bibinfo{series}{Leibniz International
  Proceedings in Informatics (LIPIcs)}}~\bibinfo{volume}{35},
  \bibinfo{publisher}{Schloss Dagstuhl--Leibniz-Zentrum fuer Informatik},
  \bibinfo{address}{Dagstuhl, Germany}, pp. \bibinfo{pages}{86--100},
  \doi{10.4230/LIPIcs.CALCO.2015.86}.
\newblock \urlprefix\url{http://drops.dagstuhl.de/opus/volltexte/2015/5528}.

\bibitemdeclare{inproceedings}{Germane:2017:PEA:3009837.3009899}
\bibitem{Germane:2017:PEA:3009837.3009899}
\bibinfo{author}{Kimball \surnamestart Germane\surnameend} \&
  \bibinfo{author}{Matthew \surnamestart Might\surnameend}
  (\bibinfo{year}{2017}): \emph{\bibinfo{title}{A Posteriori Environment
  Analysis with Pushdown Delta CFA}}.
\newblock In: {\sl \bibinfo{booktitle}{Proceedings of the 44th ACM SIGPLAN
  Symposium on Principles of Programming Languages}}, \bibinfo{publisher}{ACM},
  \bibinfo{address}{New York, NY, USA}.

\bibitemdeclare{article}{DBLP:journals/scp/GreweEPRM18}
\bibitem{DBLP:journals/scp/GreweEPRM18}
\bibinfo{author}{Sylvia \surnamestart Grewe\surnameend},
  \bibinfo{author}{Sebastian \surnamestart Erdweg\surnameend},
  \bibinfo{author}{Andr{\'{e}} \surnamestart Pacak\surnameend},
  \bibinfo{author}{Michael \surnamestart Raulf\surnameend} \&
  \bibinfo{author}{Mira \surnamestart Mezini\surnameend}
  (\bibinfo{year}{2018}): \emph{\bibinfo{title}{Exploration of language
  specifications by compilation to first-order logic}}.
\newblock {\sl \bibinfo{journal}{Sci. Comput. Program.}} \bibinfo{volume}{155},
  pp. \bibinfo{pages}{146--172}, \doi{10.1016/j.scico.2017.08.001}.
\newblock \urlprefix\url{https://doi.org/10.1016/j.scico.2017.08.001}.

\bibitemdeclare{book}{indexed}
\bibitem{indexed}
\bibinfo{author}{John~E. \surnamestart Hopcroft\surnameend},
  \bibinfo{author}{Rajeev \surnamestart Motwani\surnameend} \&
  \bibinfo{author}{Jeffrey~D. \surnamestart Ullman\surnameend}
  (\bibinfo{year}{2006}): \emph{\bibinfo{title}{Introduction to Automata
  Theory, Languages, and Computation (3rd Edition)}}.
\newblock \bibinfo{publisher}{Addison-Wesley Longman Publishing Co., Inc.},
  \bibinfo{address}{Boston, MA, USA}.

\bibitemdeclare{inproceedings}{multistage}
\bibitem{multistage}
\bibinfo{author}{Jun \surnamestart Inoue\surnameend} \& \bibinfo{author}{Walid
  \surnamestart Taha\surnameend} (\bibinfo{year}{2012}):
  \emph{\bibinfo{title}{Reasoning About Multi-stage Programs}}.
\newblock In \bibinfo{editor}{Seidl}  \cite{esop2012}.

\bibitemdeclare{techreport}{Ion:01:MML}
\bibitem{Ion:01:MML}
\bibinfo{author}{Patrick D~F \surnamestart Ion\surnameend},
  \bibinfo{author}{Nico \surnamestart Poppelier\surnameend},
  \bibinfo{author}{David \surnamestart Carlisle\surnameend} \&
  \bibinfo{author}{Robert~R \surnamestart Miner\surnameend}
  (\bibinfo{year}{2001}): \emph{\bibinfo{title}{Mathematical Markup Language
  ({MathML}) Version 2.0}}.
\newblock \bibinfo{type}{{W3C} Recommendation}, \bibinfo{institution}{W3C}.
\newblock \bibinfo{note}{Https://www.w3.org/TR/MathML2/chapter3.html}.

\bibitemdeclare{techreport}{ISO26300}
\bibitem{ISO26300}
 (\bibinfo{year}{2015}): \emph{\bibinfo{title}{{Information technology -- Open
  Document Format for Office Applications ({OpenDocument}) v1.2 -- Part 1:
  OpenDocument Schema}}}.
\newblock \bibinfo{type}{Standard}, \bibinfo{institution}{International
  Organization for Standardization}, \bibinfo{address}{Geneva, CH}.

\bibitemdeclare{incollection}{computerising}
\bibitem{computerising}
\bibinfo{author}{Fairouz \surnamestart Kamareddine\surnameend},
  \bibinfo{author}{Joe \surnamestart Wells\surnameend},
  \bibinfo{author}{Christoph \surnamestart Zengler\surnameend} \&
  \bibinfo{author}{Henk \surnamestart Barendregt\surnameend}
  (\bibinfo{year}{2014}): \emph{\bibinfo{title}{Computerising Mathematical
  Text}}.
\newblock In \bibinfo{editor}{Jörg~H. \surnamestart Siekmann\surnameend},
  editor: {\sl \bibinfo{booktitle}{Computational Logic}}, {\sl
  \bibinfo{series}{Handbook of the History of Logic}}~\bibinfo{volume}{9},
  \bibinfo{publisher}{North-Holland}, pp. \bibinfo{pages}{343 -- 396},
  \doi{https://doi.org/10.1016/B978-0-444-51624-4.50008-3}.
\newblock
  \urlprefix\url{http://www.sciencedirect.com/science/article/pii/B9780444516244500083}.

\bibitemdeclare{book}{TeXbook}
\bibitem{TeXbook}
\bibinfo{author}{Donald~E. \surnamestart Knuth\surnameend}
  (\bibinfo{year}{1986}): \emph{\bibinfo{title}{The TeXbook}}.
\newblock \bibinfo{publisher}{Addison-Wesley Professional}.

\bibitemdeclare{inproceedings}{inftyop}
\bibitem{inftyop}
\bibinfo{author}{Luis~Fdo. \surnamestart Llana~D{\'i}az\surnameend} \&
  \bibinfo{author}{Manuel \surnamestart N{\'u}{\~{n}}ez\surnameend}
  (\bibinfo{year}{1997}): \emph{\bibinfo{title}{Testing semantics for unbounded
  nondeterminism}}.
\newblock In \bibinfo{editor}{Christian \surnamestart Lengauer\surnameend},
  \bibinfo{editor}{Martin \surnamestart Griebl\surnameend} \&
  \bibinfo{editor}{Sergei \surnamestart Gorlatch\surnameend}, editors: {\sl
  \bibinfo{booktitle}{Euro-Par'97 Parallel Processing}},
  \bibinfo{publisher}{Springer Berlin Heidelberg}, \bibinfo{address}{Berlin,
  Heidelberg}, pp. \bibinfo{pages}{538--545}.

\bibitemdeclare{book}{Ordinal}
\bibitem{Ordinal}
\bibinfo{author}{Yiannis \surnamestart Moschovakis\surnameend}
  (\bibinfo{year}{1994}): \emph{\bibinfo{title}{Notes on Set Theory}},
  \bibinfo{edition}{1} edition.
\newblock {\sl \bibinfo{series}{0172-6056}}
  \bibinfo{volume}{978-1-4757-4153-7}, \bibinfo{publisher}{Springer-Verlag New
  York}, \doi{10.1007/978-1-4757-4153-7}.

\bibitemdeclare{article}{Neumann}
\bibitem{Neumann}
\bibinfo{author}{John \surnamestart von Neumann\surnameend}
  (\bibinfo{year}{1923}): \emph{\bibinfo{title}{{Zur Einführung der
  transfiniten Zahlen}}}.
\newblock {\sl \bibinfo{journal}{Acta Scientiarum Mathematicarum (Szeged)}}
  \bibinfo{volume}{1}(\bibinfo{number}{4}), pp. \bibinfo{pages}{199--208}.
\newblock
  \urlprefix\url{http://acta.fyx.hu/acta/showCustomerArticle.action?id=4981&dataObjectType=article}.
\newblock \bibinfo{note}{Auf Englisch nachgedruckt in
  GlossarWiki:Heijenoort:2002}.

\bibitemdeclare{book}{Pierce:2002:TPL:509043}
\bibitem{Pierce:2002:TPL:509043}
\bibinfo{author}{Benjamin~C. \surnamestart Pierce\surnameend}
  (\bibinfo{year}{2002}): \emph{\bibinfo{title}{Types and Programming
  Languages}}, \bibinfo{edition}{1st} edition.
\newblock \bibinfo{publisher}{The MIT Press}.

\bibitemdeclare{inproceedings}{8005074}
\bibitem{8005074}
\bibinfo{author}{V.~\surnamestart Rahli\surnameend},
  \bibinfo{author}{M.~\surnamestart Bickford\surnameend} \&
  \bibinfo{author}{R.~L. \surnamestart Constable\surnameend}
  (\bibinfo{year}{2017}): \emph{\bibinfo{title}{Bar induction: The good, the
  bad, and the ugly}}.
\newblock In: {\sl \bibinfo{booktitle}{2017 32nd Annual ACM/IEEE Symposium on
  Logic in Computer Science (LICS)}}, pp. \bibinfo{pages}{1--12},
  \doi{10.1109/LICS.2017.8005074}.

\bibitemdeclare{book}{Reynolds:2009:TPL:1611353}
\bibitem{Reynolds:2009:TPL:1611353}
\bibinfo{author}{John~C. \surnamestart Reynolds\surnameend}
  (\bibinfo{year}{2009}): \emph{\bibinfo{title}{Theories of Programming
  Languages}}, \bibinfo{edition}{1st} edition.
\newblock \bibinfo{publisher}{Cambridge University Press},
  \bibinfo{address}{New York, NY, USA}.

\bibitemdeclare{proceedings}{esop2012}
\bibitem{esop2012}
\bibinfo{editor}{Helmut \surnamestart Seidl\surnameend}, editor
  (\bibinfo{year}{2012}): \emph{\bibinfo{title}{Programming Languages and
  Systems}}. \bibinfo{publisher}{Springer}.

\bibitemdeclare{article}{Sewell:2007:OET:1291220.1291155}
\bibitem{Sewell:2007:OET:1291220.1291155}
\bibinfo{author}{Peter \surnamestart Sewell\surnameend},
  \bibinfo{author}{Francesco~Zappa \surnamestart Nardelli\surnameend},
  \bibinfo{author}{Scott \surnamestart Owens\surnameend},
  \bibinfo{author}{Gilles \surnamestart Peskine\surnameend},
  \bibinfo{author}{Thomas \surnamestart Ridge\surnameend},
  \bibinfo{author}{Susmit \surnamestart Sarkar\surnameend} \&
  \bibinfo{author}{Rok \surnamestart Strni\v{s}a\surnameend}
  (\bibinfo{year}{2007}): \emph{\bibinfo{title}{Ott: Effective Tool Support for
  the Working Semanticist}}.
\newblock {\sl \bibinfo{journal}{SIGPLAN Not.}}
  \bibinfo{volume}{42}(\bibinfo{number}{9}), pp. \bibinfo{pages}{1--12},
  \doi{10.1145/1291220.1291155}.
\newblock \urlprefix\url{http://doi.acm.org/10.1145/1291220.1291155}.

\bibitemdeclare{inproceedings}{Steele:2017:TNO:3018743.3018773}
\bibitem{Steele:2017:TNO:3018743.3018773}
\bibinfo{author}{Guy~L. \surnamestart Steele\surnameend, Jr.}
  (\bibinfo{year}{2017}): \emph{\bibinfo{title}{It's Time for a New Old
  Language}}.
\newblock In: {\sl \bibinfo{booktitle}{Proceedings of the 22Nd ACM SIGPLAN
  Symposium on Principles and Practice of Parallel Programming}},
  \bibinfo{series}{PPoPP '17}, \bibinfo{publisher}{ACM}, \bibinfo{address}{New
  York, NY, USA}, pp. \bibinfo{pages}{1--1}, \doi{10.1145/3018743.3018773}.
\newblock \urlprefix\url{http://doi.acm.org/10.1145/3018743.3018773}.

\bibitemdeclare{article}{tarski1955}
\bibitem{tarski1955}
\bibinfo{author}{Alfred \surnamestart Tarski\surnameend}
  (\bibinfo{year}{1955}): \emph{\bibinfo{title}{A lattice-theoretical fixpoint
  theorem and its applications.}}
\newblock {\sl \bibinfo{journal}{Pacific J. Math.}}
  \bibinfo{volume}{5}(\bibinfo{number}{2}), pp. \bibinfo{pages}{285--309}.
\newblock \urlprefix\url{https://projecteuclid.org:443/euclid.pjm/1103044538}.

\bibitemdeclare{inproceedings}{metalambda}
\bibitem{metalambda}
\bibinfo{author}{Kazunori \surnamestart Tobisawa\surnameend}
  (\bibinfo{year}{2015}): \emph{\bibinfo{title}{A Meta Lambda Calculus with
  Cross-Level Computation}}.
\newblock In: {\sl \bibinfo{booktitle}{POPL '15}}, pp.
  \bibinfo{pages}{383--393}.

\bibitemdeclare{inproceedings}{lambdazfc}
\bibitem{lambdazfc}
\bibinfo{author}{Neil \surnamestart Toronto\surnameend} \& \bibinfo{author}{Jay
  \surnamestart McCarthy\surnameend} (\bibinfo{year}{2012}):
  \emph{\bibinfo{title}{Computing in Cantor's Paradise with $\lambda$ ZFC}}.
\newblock In \bibinfo{editor}{Tom \surnamestart Schrijvers\surnameend} \&
  \bibinfo{editor}{Peter \surnamestart Thiemann\surnameend}, editors: {\sl
  \bibinfo{booktitle}{Functional and Logic Programming}},
  \bibinfo{publisher}{Springer Berlin Heidelberg}, \bibinfo{address}{Berlin,
  Heidelberg}, pp. \bibinfo{pages}{290--306}.

\bibitemdeclare{article}{Wiener}
\bibitem{Wiener}
\bibinfo{author}{Norbert \surnamestart Wiener\surnameend}
  (\bibinfo{year}{1914}): \emph{\bibinfo{title}{A Simplification of the Logic
  of Relations}}.
\newblock {\sl \bibinfo{journal}{Proceedings of Cambridge Philosophical
  Society}} \bibinfo{volume}{17}, pp. \bibinfo{pages}{387--390}.
\newblock \bibinfo{note}{Nachgedruckt in GlossarWiki:Heijenoort:2002}.

\end{thebibliography}
 
\appendix

\note{}

\section{Basic Logic and Mathematics}

This appendix gives a brief overview of some concepts which are common enough in mathematics, but which are often represented in different ways, to say how they are used in this paper.

\subsection{Metavariable Conventions}
\label{sec:meta-decl}
For this section,  $\MetaVar$ stands for an arbitrary
metavariable (a meta-metavariable).
Statements of the form ``let $\MetaVar$ range over $\mathcal{C}$''
declare and define $\MetaVar$ as a
metavariable that stands for some element of the class
$\mathcal{C}$.
\note{}

We use single letters (either Roman or Greek) for metavariables.
\note{}

Whenever we declare a metavariable $\MetaVar$ as ranging over a class,
this also defines as ranging over that class all variants of
$\MetaVar$ obtained by either
(1)~adding a subscript $i \in \NatSet$ to $\MetaVar$ to produce
    $\MetaVar_i$ (e.g., $\MetaVar_0$, $\MetaVar_1$, $\MetaVar_2$, etc),
(2)~adding a single, double, or triple prime to $\MetaVar$,
    producing respectively in $\MetaVar'$, $\MetaVar''$, and
    $\MetaVar'''$, or
(3)~a combination of (1) and~(2).

In contrast, we use superscripts (e.g., $\MetaVar^1$, $\MetaVar^2$)
and accents (e.g., $\Bar{\MetaVar}$, $\tilde{\MetaVar}$) to
distinguish metavariables that are in some way related to
the corresponding undecorated metavariable,
but not necessarily ranging over the same class.
For example, if we have declared $\MetaVar$ to range over the set
$\SetS$, we might have
$\MetaVar^0$ ranging over $\SetS^0$,
$\MetaVar^1$ ranging over $\SetS^1$,
and $\SetS^1 \subset \SetS^0 \subset \SetS$.

\note{}

\subsection{Sets}

The mathematical foundation we use is set theory with choice.
ZFC is suitable, so are other
variants.
If $P(\StuffX)$ is a proposition of first-order logic that mentions
$\Stuff$, then
(1) $P(\StuffY)$ differs from $P(\StuffX)$ only by mentioning
    $\StuffY$ instead of $\StuffX$, and
(2) the notation $\AllSuchThat{\StuffX}{P(\StuffX)}$
stands for $\AllSuchThat{\StuffX\in\SetS}{P(\StuffX)}$ for
some set $\SetS$ which is left to the reader to infer from the context
of discussion.
Given some expression $f(\Stuff_1,\ldots,\Stuff_n)$ mentioning
variables $\Stuff_1$, $\ldots\,$, $\Stuff_n$, we use the notation
$\AllSuchThat{f(\Stuff_1,\ldots,\Stuff_n)}{P(\Stuff_1,\ldots,\Stuff_n)}$
for

\noindent $
  \AllSuchThat{\StuffY}
    {\exists{\Stuff_1,\ldots,\StuffX_n}.\;
       \StuffY=f(\StuffX_1,\ldots,\StuffX_n)\wedge{P(\StuffX_1,\ldots,\StuffX_n)}}
$.
Given two sets $\StuffX$ and $\StuffY$ we use the notation $\StuffX\mathrel{\bot}\StuffY$ to mean `$\StuffX$ and $\StuffY$ are disjoint.'

\subsection{Pairs}

We rely on a operator $\Pair{\,\cdot\,}{\,\cdot\,}$ for building
\DefiningUse{ordered pairs} and corresponding \DefiningUse{projection}
operators $\FstSym$ and $\SndSym$, such that
if $\StuffZ = \Pair{\StuffX}{\StuffY}$,
then $\Fst{\StuffZ} = \StuffX$ and $\Snd{\StuffZ} = \StuffY$.
We require that it is impossible for a pair to also be a set of pairs
and that the natural numbers do not overlap with pairs.%
\footnote{%
  We therefore can not use Kuratowski's encoding of pairs where
  $\Pair{\StuffX}{\StuffY}=\Set{\Set{\StuffX},\Set{\StuffX,\StuffY}}$,
  because (for example)
  $\Set{\Pair{\StuffX}{\StuffX}}=\Set{\Set{\Set{\StuffX}}}=\Pair{\Set{\StuffX}}{\Set{\StuffX}}$.
  Similarly, we can not use the ``short'' encoding where
  $\Pair{\StuffX}{\StuffY}=\Set{\StuffX,\Set{\StuffX,\StuffY}}$
  together with von Neumann's encoding
  of natural numbers (actually of all ordinal numbers) where
  $0=\EmptySet$ and $i+1=i\cup\Set{i}$ because
  \(
      \Pair{0}{0}
    = \Set{0,\Set{0,0}}
    = \Set{\EmptySet,\Set{\EmptySet,\EmptySet}}
    = \Set{\EmptySet,\Set{\EmptySet}}
    = \Set{\EmptySet}\cup\Set{\Set{\EmptySet}}
    = 1\cup\Set{1}
    = 2
  \).
  We can use Wiener's encoding of pairs where
  $\Pair{\StuffX}{\StuffY}=\Set{\Set{\Set{\StuffX},\EmptySet},\Set{\Set{\StuffY}}}$,
  because
  in this encoding a pair can not be a set of pairs, a set of sets of
  pairs, or a von Neumann ordinal number.
  We can also work in a set theory with a primitive pairing operator.%
  }
\relax
Given two sets $\SetS$ and $\SetT$, the \DefiningUse{product set}
$\SetS \times \SetT$ is the set of pairs
$\AllSuchThat{\Pair{\StuffX}{\StuffY}}{\StuffX\in{\SetS}\mbox{ and }\StuffY\in{\SetT}}$.
Let \DefiningUse{tuple} notation be defined so that
\(
  \Tuple{\StuffX_1,\StuffX_2,\StuffX_3,\ldots,\StuffX_n}
  =
  \Pair{\Tuple{\StuffX_1,\StuffX_2,\StuffX_3,\ldots,\StuffX_{n-1}}}{\StuffX_n}
\).
\note{}

\subsection{Relations}
\label{sec:prel-relations}
Let $\RelationR$
range over sets of pairs.
The statement $\Pair{\StuffX}{\StuffY}\in\RelationR$ can be written
with three
kinds of alternate notation: $\RelationR(\StuffX,\StuffY)$,
and $\StuffX\mathrel{\RelationR}\StuffY$,
and $\StuffX\GroundRelation{\RelationR}{}\StuffY$.

A relation $\RelationR$ is \DefiningUse{reflexive w.r.t.\ $\SetS$} iff
$\RelationR\supseteq\AllSuchThat{\Pair{\StuffX}{\StuffX}}{\StuffX\in\SetS}$.
As is common practice, if we mention that a relation is reflexive
without saying what set $\SetS$ this is with respect to, this means we
are leaving it to the reader to infer from the context of discussion
which set $\SetS$ to use.

Let $\RelationR^*$ be the reflexive and transitive closure of
$\RelationR$ and let $\RelationR^=$ be the reflexive, symmetric, and
transitive closure of $\RelationR$; in both cases we use the
above-mentioned convention that the reader must infer the set $\SetS$
w.r.t.\ which to take the reflexive closure.
Let $\StuffX\ReflTransRelation{\RelationR}{}\StuffY$ mean
$\StuffX\GroundRelation{\RelationR^*}{}\StuffY$, and let
$\StuffX\ReflTransSymmRelation{\RelationR}{}\StuffY$ mean
$\StuffX\GroundRelation{\RelationR^=}{}\StuffY$.

A relation is an \DefiningUse{equivalence} iff it is symmetric and
transitive.
Given an equivalence relation $\RelationR$, let
$\EquivClass{\StuffX}{\RelationR}=\AllSuchThat{\StuffY}{\Pair{\StuffX}{\StuffY}\in\RelationR}$
be the \DefiningUse{equivalence class of $\StuffX$ w.r.t.\ $\RelationR$} and
let $\EquivClass{\StuffX}{\RelationR}$ be an equivalence class of
$\RelationR$.%
\note{}

\note{}

A relation $\RelationR$ is \DefiningUse{terminating} iff
there is no infinite sequence
$\StuffX_1$, $\StuffX_2$, $\ldots$ such that
\(
  \StuffX_1 \GroundRelation{\RelationR}{}
    \StuffX_2 \GroundRelation{\RelationR}{}
      \cdots{}
\).
If $\StuffX \ReflTransRelation{\RelationR}{} \StuffY$, and there
exists no $\StuffZ$ such that
$\StuffY \GroundRelation{\RelationR}{} \StuffZ$, then we call $\StuffY$
\DefiningUse{an $\RelationR$-normal form of $\StuffX$}.
If $\RelationR$ is terminating, then it can be used for
\DefiningUse{induction}: If it can be shown that $\RelationR$ is
terminating and
\(
  \forall{\StuffX}\in\SetS.\;
                (\forall{\StuffY}\in\SetS.\;
                   \StuffX\GroundRelation{\RelationR}{}\StuffY\Rightarrow{P(\StuffY)})
    \Rightarrow {P(\StuffX)}
\),
then it follows that $\forall{\StuffX}\in\SetS.\;P(\StuffX)$.

A relation is \DefiningUse{a partial order} on $\SetS$ iff it is transitive and antisymmetric.
A partial order is \DefiningUse{strict} iff it is irreflexive.
A non-strict partial order, $\leq$, is a \DefiningUse{total order} on $\SetS$ iff for all $\StuffX ,\StuffY\in\SetS$ either $\StuffX\leq\StuffY$ or $\StuffY\leq\StuffX$.
A strict partial order, $<$, is a \DefiningUse{strict total order} on $\SetS$ iff for all $\StuffX ,\StuffY\in\SetS$ s.t. $\StuffX\neq\StuffY$ either $\StuffX <\StuffY$ or $\StuffY <\StuffX$.
\note{}

\subsection{Functions}

A \DefiningUse{function} is a relation $f$ such that for all
$\AnyX$, $\AnyY$, and $\AnyZ$, if
\(
  \Set{ \Pair{\AnyX}{\AnyY}, \Pair{\AnyX}{\AnyZ} }
  \subseteq f
\)
then $\AnyY=\AnyZ$.
Let
\(
    \FunSpace{\SetS}{\SetT}
  = \AllSuchThat{f}
      {             f\subseteq\SetS\times\SetT
       \mbox{ and } f\mbox{ is a function}}
\).
Let $f$ be \DefiningUse{from $\SetS$ to $\SetT$} iff
$f\in\FunSpace{\SetS}{\SetT}$.
A function $f$ is \DefiningUse{injective} iff $f^{-1}$ is a function.
If $\Pair{X}{Y}\in{f}$ for some $Y$, then $f(X)$ denotes $Y$,
otherwise $f(X)$ is undefined.
A function $f$ is  \DefiningUse{total} on $\SetS$ iff $f(X)$ is defined for all $X\in\SetS$.%
\note{}
Given a function $f$, let
\(
    \PerturbFunAt{f}{X}{Y}
  = (f\setminus\AllSuchThat{Z\in{f}}{\Fst{Z}=X})\cup\Set{\Pair{X}{Y}}
\).
\note{}

A \DefiningUse{fixed point} of a function $f$ is some $x$ for which $f(\StuffX)=\StuffX$.
If the set of fixed points of $f$ has a greatest lower bound which is itself a fixed point, then we call this the \DefiningUse{least fixed point} of $f$
 and if it has a least upper bound which is itself a fixed point, then we call this the \DefiningUse{greatest fixed point} of $f$.

A function is $f$ \DefiningUse{order preserving} w.r.t a partial ordering $leq$ if $f(\StuffX )\leq f(\StuffY )$ iff $\StuffX\leq\StuffY$.

\subsection{Sequences}

Given a set $\SetS$ which is not a relation (if $\SetS$ contains only
pairs then instead the notation refers to the definition of
$\RelationR^*$ from \autoref{sec:prel-relations}, the reflexive and
transitive closure of $\RelationR$), let
$\SeqSet{\SetS}$, the
set of \DefiningUse{finite sequences of elements in $\SetS$}, be the
set of all finite functions $f$ such that $\Range{f}\subseteq\SetS$,
and $\Domain{f}\subseteq\NatSet$, and $m<n\in\Domain{f}$ implies
$m\in\Domain{f}$.
\note{}

\begin{convention}[Metavariables over Sequences]
  \label{conv:sequence-metavariables}
If $\MetaVar$ is declared to range over $\SetS$, then
$\SeqVar{\MetaVar}$ is automatically declared to range over
$\SeqSet{\SetS}$.
\end{convention}

The notation $\Seq{\MetaVar_0, \ldots, \MetaVar_n}$ stands for the
least-defined function $\SeqVar{\MetaVar}$ such that
$\SeqVar{\MetaVar}(i) = \MetaVar_i$ for all $i \in \Set{0, \ldots, n}$.
For example, the singleton sequence $\Seq{\MetaVar}$ containing
$\MetaVar$ as its only element is $\Set{\Pair{0}{\MetaVar}}$, and
we have
\(
  \Seq{\MetaVar_0, \MetaVar_1, \MetaVar_2}
  =
  \Set{\Pair{0}{\MetaVar_0},
       \Pair{1}{\MetaVar_1},
       \Pair{2}{\MetaVar_2}}
\).
The \DefiningUse{component of a sequence $\SeqVar{\MetaVar}$ at index
$i$} is simply $\Nth{\SeqVar{\MetaVar}}{i}$.
Note that the first component of a sequence is at index $0$, and that
the \DefiningUse{empty sequence} $\SeqEmpty$ is merely the empty set.
The \DefiningUse{length} of a sequence $\SeqVar{\MetaVar}$ is the
smallest $n\in\NatSet$ which is larger than all elements of
$\Domain{\SeqVar{\MetaVar}}$.
The \DefiningUse{concatenation} of sequences $\SeqVar{\MetaVar}_1$ and
$\SeqVar{\MetaVar}_2$ is
\(
  \SeqConcat{\SeqVar{\MetaVar}_1}{\SeqVar{\MetaVar}_2}
  =
  \SeqVar{\MetaVar}_1
    \cup
  \AllSuchThat
    {\Pair{\SeqLen{\SeqVar{\MetaVar}_1} + i}{\MetaVar}}
    {\Pair{i}{\MetaVar} \in \SeqVar{\MetaVar}_2}
\).

Note that $\Triple{\SeqSet{\SetS}}{\SeqConcatSym}{\SeqEmpty}$
forms a monoid, i.e., the following equalities hold:
$$
\begin{array}{@{}c@{\qquad}c@{\qquad}c@{}}
\SeqConcat{\SeqEmpty}{\SeqVar{\MetaVar}} = \SeqVar{\MetaVar} &
\SeqConcat{\SeqVar{\MetaVar}}{\SeqEmpty} = \SeqVar{\MetaVar} &
\SeqConcat{\SeqConcatP{\SeqVar{\MetaVar}_1}{\SeqVar{\MetaVar}_2}}{\SeqVar{\MetaVar}_3} =
\SeqConcat{\SeqVar{\MetaVar}_1}{\SeqConcatP{\SeqVar{\MetaVar}_2}{\SeqVar{\MetaVar}_3}}
\end{array}\eqno\EEM
$$

\note{}
  
\note{}

\section{Proof of Key Results}
Section 2 is sufficient for any reader who wants an outline of our definition in order to interpret those pieces of MBNF it defines.
This appendix will be of interest either to those readers who are looking to extend our definition to define more uses of MBNF, or to those readers who want to reassure themselves that the sort of entities described in section 2.1 can always be thought to exist.
While it is our intention that our definition should be easy to work with, a deeper working knowledge of set theory is assumed for this appendix than the rest of this document and everything apart from section 3 may be read without this section.

For this appendix we use Wiener's \cite{Wiener} encoding of pairs and von Neumann's encoding of ordinals, the natural numbers \cite{Neumann} and cardinal assignment \cite{Ordinal}.
We also make use of the axiom of choice.
\begin{lemma}
\label{lem:tree-sizes} Given a countable set $A$ and a set $B$  of all trees $C$ such that the interior nodes of $C$ are elements of $A$ and the leaf nodes of $C$ are elements of $\omega_1$, $|B|=\aleph_1$.
\begin{proof}
Let $D$ be the set of all trees $C$ such that every element of $C$ is an element of $\omega_1$.
Since we can make a bijection between members of $A$ to a members of $\omega$ and the function $f\in\FunSpace{\omega_1}{\omega_1}$ s.t. $f(x)=\omega +x$ is a bijection, $|C|=|D|$.
For a finite subset, $\SetS$, of $\omega_1$ the relation $<$ on $\SetS$ is a finite subset of $\omega_1\times\omega_1$.
Assuming choice, $|\omega_1\times\omega_1|=\aleph_1$.
The cardinality of the set of finite subsets of $\omega_1$ is  $\aleph_1$. 
So $|C|=|D|=\aleph_1$.
\end{proof}
\end{lemma}

We can now go on to show that $\ObjectSet$ and $\ArrangementSet$ are well defined. 
We do so by producing a model within set theory that fulfils most of the constraints in section 2.1., which are written out formally in Appendix 
D.\footnote{In order to simplify the proof this model allows the use of arrangements consisting of a single pointer in the formation of objects. 
It is not too difficult to rule this case out.
}
This proof requires that the reader pick some appropriate values for $\SymbolSet$, $\PositionSet$, $\PointerSet$, $\ContextHole$, $\epsilon$ and $\BarSign$.
The definition of these sets in Appendix D is adequate.

We for a given ordinal $i$ we define  a set of tuples $\mtnormsf{OPAC}_i$ which may be thought of as getting closer to the tuple $\Tuple{\ObjectSet,\mtnormsf{ptr},\ArrangementSet,\mtnormsf{Core}}$.

\begin{definition}[OPAC]
\noindent

\noindent {\bf 0 Case:}

Let $\mtnormsf{Obj}_0=\{\ContextHole\}$
 
Let $\mtnormsf{ptrSpace}_0=\AllSuchThat{x\in\FunSpace{\mtnormsf{Obj}_0}{\PointerSet}}{x\text{ is total on }\mtnormsf{Obj}_0\wedge x\text{ is injective}}$.

 Let $$\begin{array}{rcl}
 \mtnormsf{OPAC}_0=\{\Tuple{\mtnormsf{Obj}_0,\mtnormsf{ptr}_0,\mtnormsf{Arr}_0,\mtnormsf{Core}_0}&|&\mtnormsf{ptr}_0\in\mtnormsf{ptrSpace}_0\wedge
 \mtnormsf{Core}_0=\mtnormsf{Arr}_0\setminus\{\epsilon\}\\
 &&\wedge\,\mtnormsf{Arr}_0=\NatSet\cup\{\epsilon\}
 \cup\SymbolSet\cup
 \mtnormsf{ptr}_0(x)\}
\end{array}  
 $$
 
 \noindent {\bf +1 Case:}
 
For $\Tuple{\mtnormsf{Obj}_n^k,\mtnormsf{ptr}_n^k,\mtnormsf{Arr}_n^k,\mtnormsf{Core}_n^k}\in \mtnormsf{OPAC}_n$

Let $\mtnormsf{Accent}_{n+1}^{k}=(\{\BarSign\}\times\SymbolSet)\times(\mtnormsf{Arr}_n^{k}
 \setminus\{\epsilon\})$.

Let $\mtnormsf{Core}_{n+1}^k=\mtnormsf{Core}_n^k\cup\mtnormsf{Accent}_{n+1}^k$.
 
Let $\mtnormsf{Layout}_{n+1}^k = \mtnormsf{Arr}_{n}^k \times\AllSuchThat{ x\in\FunSpace{\PositionSet}{\mtnormsf{Arr}_{n}^k\setminus\{\epsilon\}}}{x\neq\emptyset}$.

 Let $\mtnormsf{Seq}_{n+1}^k=\mtnormsf{Arr}_{n}^k\times
 \mtnormsf{CoreArr}_{n+1}^k$.

  Let  $\mtnormsf{Obj}_{n+1}^k=\mtnormsf{Obj}_n^k\cup
  \AllSuchThat{x\in\mathcal{P}(\mtnormsf{Arr}_n^k)}{|x\,|\leq \aleph_0}$
  
 Let $$\begin{array}{rcl}
 \mtnormsf{ptrSpace}_{n+1}=\{x\in
 \FunSpace{\mtnormsf{Obj}_{n+1}^k}{\PointerSet}&|&\Tuple{\mtnormsf{Obj}_{n},\mtnormsf{ptr}_{n},\mtnormsf{Arr}_{n},\mtnormsf{Core}_{n}}\in\mtnormsf{OPAC}_{n}\wedge \\
 &&x\text{ is total on }\mtnormsf{Obj}_{n+1}^k\wedge x\text{ is injective}\}
 \end{array}$$
 
  Let $\mtnormsf{ptr}_{n+1}^{k,i}(x)\in\mtnormsf{ptrSpace}_{n+1}$ such that $\mtnormsf{ptr}_{n+1}^{k,i}(x)\subseteq\mtnormsf{ptr}_{n+1}^{k}$
  if such a set exists.
  If no such set exists let $\mtnormsf{ptr}_{n+1}^{k,0}=\emptyset$
  
Let $  
  \mtnormsf{Arr}_{n+1}^{k,i}=\mtnormsf{Arr}_n^k\cup
  \mtnormsf{CoreArr}_{n+1}^k
  \cup\mtnormsf{Layout}_{n+1}^k\cup\mtnormsf{Seq}_{n+1}^k\cup
   \AllSuchThat{\mtnormsf{ptr}_{n+1}^{k,i}(x)}{x\in\mtnormsf{Obj}_{n+1}}$  
   if $\mtnormsf{ptr}_{n+1}^{k,i}(x)$ is defined and $\emptyset$ otherwise.
  
  Let $$\begin{array}{rcl}
  \mtnormsf{OPAC}_{n+1}=\{\Tuple{\mtnormsf{Obj}_{n+1}^k,\mtnormsf{ptr}_{n+1}^{k,i},\mtnormsf{Arr}_{n+1}^{k,i},\mtnormsf{Core}_{n+1}^k}&|&\Tuple{\mtnormsf{Obj}_{n},\mtnormsf{ptr}_{n},\mtnormsf{Arr}_{n},\mtnormsf{Core}_{n}}\in\mtnormsf{OPAC}_{n}\\
 && \wedge \,\mtnormsf{Arr}_{n+1}^{k,i}\neq\emptyset\}
  \end{array}
$$

  \noindent {\bf Limit Case:}
  
  We now define the above functions for a limit point $\varepsilon$.

Let $$
\begin{array}{rcl}
\mtnormsf{stack}=\{\SetS\subseteq\bigcup\limits_{i<\varepsilon} \mtnormsf{OPAC}_i&|&
(\Tuple{\mtnormsf{Obj}_i^k,\mtnormsf{ptr}_i^k,\mtnormsf{Arr}_i^k,\mtnormsf{Core}_i^k}\in \SetS
\wedge\Tuple{\mtnormsf{Obj}_j^l,\mtnormsf{ptr}_j^l,\mtnormsf{Arr}_j^l,\mtnormsf{Core}_j^l}\in \SetS)\\
&&\begin{array}{ll}
\Rightarrow&
((j<i\Rightarrow(\mtnormsf{Obj}_j^l\subseteq\mtnormsf{Obj}_i^k
\wedge \mtnormsf{Arr}_j^l\subseteq\mtnormsf{Arr}_i^k\wedge \mtnormsf{ptr}_j^l\subseteq \mtnormsf{ptr}_i^k))
\\
&\begin{array}{ll}
\wedge&(j<i\vee i<j\\
&\begin{array}{ll}
\vee&\Tuple{\mtnormsf{Obj}_i^k,\mtnormsf{ptr}_i^k,\mtnormsf{Arr}_i^k,\mtnormsf{Core}_i^k}
=\Tuple{\mtnormsf{Obj}_j^l,\mtnormsf{ptr}_j^l,\mtnormsf{Arr}_j^l,\mtnormsf{Core}_j^l})
\end{array}\\
\wedge&\forall n<\varepsilon,\, \Tuple{\mtnormsf{Obj}_n^m,\mtnormsf{ptr}_n^m,\mtnormsf{Arr}_n^m,\mtnormsf{Core}_n^m}\in\SetS )\}
\end{array}
\end{array}
\end{array}
$$

For $\SetS \in\mtnormsf{stack}$

Let $$\mtnormsf{Obj}_\varepsilon^\SetS =(\bigcup\AllSuchThat{\mtnormsf{Obj}_i}{\Tuple{\mtnormsf{Obj}_i^k,x,y,z}\in \SetS})
\cup\AllSuchThat{x\in\mathcal{P}(\bigcup\AllSuchThat{\mtnormsf{Arr}_i^k}{\Tuple{a,b,\mtnormsf{Arr}_i^k,c}\in \SetS})}{|x\,|\leq \aleph_0}
$$

 Let $\mtnormsf{ptr}_\varepsilon^{\SetS,k}$ be a bijection between $\SetS_{\varepsilon}\subseteq\PointerSet$ and $\mtnormsf{Obj}_{\varepsilon}$ such that for all $\Tuple{\mtnormsf{Obj}_i,\mtnormsf{ptr}_i,\mtnormsf{Arr}_i,\mtnormsf{Core}_i}\in \SetS$, $\mtnormsf{ptr}_i \subseteq\mtnormsf{ptr}_\varepsilon^{\SetS,k}$.
 If no such bijection exists, let $\mtnormsf{ptr}_\varepsilon^{\SetS,0}=\emptyset$.

Let $\mtnormsf{Arr}_\varepsilon^{\SetS,k} =
\AllSuchThat{\mtnormsf{ptr}_\varepsilon^{\SetS,k}(x)}{x\in\mtnormsf{Obj}_{\varepsilon}^\SetS}
\cup\bigcup\AllSuchThat{\mtnormsf{Arr}_i}{\Tuple{a,b,\mtnormsf{Arr}_i^k,c}\in \SetS}$ if  $\mtnormsf{ptr}_\varepsilon^{\SetS,k}(x)$ is defined and $\emptyset$ otherwise.

Let $\mtnormsf{Core}_{\varepsilon}^\SetS=
\bigcup\AllSuchThat{\mtnormsf{CoreArr}_{i}}{{\Tuple{x,y,z,\mtnormsf{Core}_i^k}\in \SetS}}$.

  Let $$
  \mtnormsf{OPAC}_{\varepsilon}=\AllSuchThat{\Tuple{
  \mtnormsf{Obj}_{\varepsilon}^\SetS,\mtnormsf{ptr}_{\varepsilon}^{\SetS,k},\mtnormsf{Arr}_{\varepsilon}^{\SetS,k},\mtnormsf{Core}_{\varepsilon}^\SetS}}{\SetS \in\mtnormsf{stack} \wedge \,\mtnormsf{Arr}_{n+1}^{k,i}\neq\emptyset}
$$
\end{definition}

\begin{lemma}$\mtnormsf{OPAC}_i$ is Non-Empty for all Ordinals $i$.
\begin{proof}
The only way $\mtnormsf{OPAC}_i$ may be empty for some $i$ is if $|\mtnormsf{Obj}_i^k|>\PointerSet$ for some $\mtnormsf{Obj}_i^k$ such that $\Tuple{\mtnormsf{Obj}_i^k,a,b,c}\in\mtnormsf{OPAC}_i$. 
We prove by induction on the size of $\mtnormsf{Obj}_n^k$ and $\mtnormsf{Arr}_n^k$ such that $\Tuple{\mtnormsf{Obj}_n^k,x,\mtnormsf{Arr}_n^k,y}\in \mtnormsf{OPAC}_n$  that this cannot be the case.

\noindent {\bf 0 Case:}

$|\mtnormsf{Obj}_0 |=1\leq\aleph_1$ and for all $\mtnormsf{Arr}_0 \in \mtnormsf{ArrSpace}_0$, $|\mtnormsf{Arr}_0 |=\aleph_0\leq\aleph_1$.

\noindent {\bf +1 Case:}

If, for all $\Tuple{\mtnormsf{Obj}_n^k,x,\mtnormsf{Arr}_n^k,y}\in \mtnormsf{OPAC}_n$, $|\mtnormsf{Obj}_n^k |\leq\aleph_1$ and $|\mtnormsf{Arr}_n^k |\leq\aleph_1$, then, for all $\mtnormsf{Obj}_{n+1}^k$, $|\mtnormsf{Obj}_{n+1}^k |\leq\aleph_1$,
  provided we have some way of ordering $\mtnormsf{Obj}_i^k$ and $\mtnormsf{Arr}_i^k$.
With choice this follows quite easily from the fact that the cardinality of the set of subsets of $\aleph_1$ which are of cardinality less than or equal to $\aleph_0$ is $(2^{\aleph_0})^{\aleph_0}=2^{\aleph_0\cdot\aleph_0}=2^{\aleph_0}$.
  As,  for all $\mtnormsf{Obj}_{n+1}^k$ there exists some $\mtnormsf{Obj}_{n}^j$ s.t. $\mtnormsf{Obj}_{n+1}^k\subseteq\mtnormsf{Obj}_{n}^j$, there exists some $\mtnormsf{ptr}$ which assigns pointers for $\mtnormsf{Obj}_{n+1}^k$ and which may also be used to assign pointers for 
  $\mtnormsf{Obj}_{n}^j$.
  
    It is easy to observe that, if $|\mtnormsf{Obj}_n^k |\leq\aleph_1$ and $|\mtnormsf{Arr}_n^k |\leq\aleph_1$, then $|\mtnormsf{Arr}_{n+1}^{k,i} |\leq\aleph_1$ since neither $\mtnormsf{Accent}_{n+1}^k$, nor $\mtnormsf{Layout}_{n+1}^k$ nor $\mtnormsf{Seq}_{n+1}^k$ can add cardinality greater than $\aleph_1$.
    
\noindent {\bf Limit Case:}

\noindent 

We show $\exists \varepsilon;\forall i<\varepsilon;\, (|\mtnormsf{Obj}_i |\leq\aleph_1\wedge|\mtnormsf{Arr}_i |\leq\aleph_1)\Rightarrow (|\mtnormsf{Obj}_\varepsilon |\leq\aleph_1\wedge|\mtnormsf{Arr}_\varepsilon |\leq\aleph_1).$
 We note that no $\mtnormsf{Arr}_i$ has $\aleph_0$ sub-arrangements and all such $\mtnormsf{Arr}_i$ can be readily identified with some finite tree whose interior nodes are labelled corresponding the operations accenting, concatenation and the finite number of possible combinations of Subscript, superscript etc.\ and whose leaf nodes are labelled with members of the set $\omega_1$.
So, by \autoref{lem:tree-sizes}, $|\bigcup\limits_{i=0}^{\varepsilon}\mtnormsf{Arr}_i |\leq\aleph_1$.
 Similarly we may readily identify each set in $\mtnormsf{Obj}_i$ apart from the $\ContextHole$ with some countable subset of the set of trees we used to define each $\mtnormsf{Arr}_i$.
 The cardinality of the countable subsets of a set of size $\aleph_1$ is $(2^{\aleph_0})^{\aleph_0}=2^{\aleph_0\cdot\aleph_0}
 =2^{\aleph_0}$.
 So $|\bigcup\limits_{i=0}^{\varepsilon}\mtnormsf{Obj}_i |\leq \aleph_1$.
 The desired result follows easily.
   As $\mtnormsf{Obj}_{\varepsilon}\subseteq\mtnormsf{Obj}_{i}$, for each $i\leq\varepsilon$ there exists some $\mtnormsf{ptr}$ which assigns pointers for $\mtnormsf{Obj}_{\varepsilon}$ and which may also be used to assign pointers for 
  $\mtnormsf{Obj}_{i}$.
\end{proof}
\end{lemma}

\begin{definition}[fun]
Let $
\StuffZ=\AllSuchThat{\mtnormsf{OPAc}_i}{i<\kappa}$ for some $\kappa<\omega_2$.
We define a function $\mtnormsf{fun}\in\FunSpace{\StuffZ}{\StuffZ}$ such that $\mtnormsf{fun}(\mtnormsf{OPAC}_i)=\mtnormsf{OPAC}_{i+1}$
\end{definition}

\begin{lemma}\mtnormsf{fun} has a least fixed point.
\begin{proof}
The  Knaster–Tarski theorem \cite{tarski1955} tells us that any any order-preserving function on a complete lattice has a least fixed point.
 For $\mtnormsf{OPAC}_a,\mtnormsf{OPAC}_b\in\StuffZ$ we define $\leq$ such that $\mtnormsf{OPAC}_a\leq\mtnormsf{OPAC}_b$ iff 
 (either,  for all $\Tuple{\mtnormsf{Obj}_b,p,q,r} \in\mtnormsf{OPAC}_b$, there exists $\Tuple{\mtnormsf{Obj}_a,b,c,d} \in\mtnormsf{OPAC}_a$ s.t. $\mtnormsf{Obj}_a\subset \mtnormsf{Obj}_b$,
  or, for all $\Tuple{\mtnormsf{Obj}_b,p,\mtnormsf{Arr}_b,r} \in\mtnormsf{OPAC}_b$, there exists $\Tuple{\mtnormsf{Obj}_a,b,\mtnormsf{Arr}_a,d} \in\mtnormsf{OPAC}_a$ s.t. ($\mtnormsf{Obj}_a= \mtnormsf{Obj}_b$ and $\mtnormsf{Arr}_a\subset \mtnormsf{Arr}_b$)).
 $\StuffZ$ is a complete lattice ordered by $\leq$ and $\mtnormsf{fun}$ is an order preserving function on $\StuffZ$.\footnote{Note that the way in which we have defined $\mtnormsf{Obj}$ and $\mtnormsf{Arr}$ is such that $j\leq i$ implies 
 ($\mtnormsf{Obj}_j\subseteq \mtnormsf{Obj}_i$ and $\mtnormsf{Obj}_j\subseteq \mtnormsf{Obj}_i$). 
 Note also that, for all limit ordinals $\varepsilon\leq\kappa$; $\Pair{\mtnormsf{Obj}_\varepsilon}{\mtnormsf{Arr}_\varepsilon}\in\StuffZ$.
 Finally note that $\omega_2$ is large enough that it has a larger cardinality than any $\mtnormsf{Obj}_i$ or $\mtnormsf{Arr}_i$, so we can select some $\kappa$ larger than the partition of $\mtnormsf{Obj}_i$ and $\mtnormsf{Arr}_i$ into the extra elements added at each stage.}
\end{proof}
\end{lemma}

\begin{theorem}
$\ObjectSet$ and $\ArrangementSet$ are well defined.
\begin{proof}
For some tuple $a$ in $\mtnormsf{OPAC}_i$ there exists some tuple $b$ in $\mtnormsf{OPAC}_i$ such that:\begin{enumerate}
\item The first member of $b$ contains all the objects our rules say must exist if $\ArrangementSet$ is at least the third member of $a$,
\item The third member of $b$ contains all the arrangements that our rules say must exist if $\ArrangementSet$ is at least the third member of $a$, $\ObjectSet$ is at least the first member of $a$ and $\mtnormsf{ptr}$ is at least the second member of $a$.
\end{enumerate}
If $\mtnormsf{OPAC}_i$ is the least fixed point of $\mtnormsf{fun}$ then $\mtnormsf{OPAC}_{i+1}=\mtnormsf{OPAC}_i$.

We now take the least fixed point, $\mtnormsf{lfp}(\mtnormsf{fun})$, of $\mtnormsf{fun}\in\FunSpace{\StuffZ}{\StuffZ}$ and select some tuple $\Tuple{\mtnormsf{Obj},\mtnormsf{ptr},\mtnormsf{Arr},\mtnormsf{Core}}\in\mtnormsf{lfp}(\mtnormsf{fun})$.
The first member of the tuple gives us a model for $\ObjectSet$ and the third member, $\ArrangementSet$.
\end{proof}
\end{theorem}

\section{Examples of our Definition in Action}

\subsection{Call by Need}
The following example is derived from Chang and Felleisen \cite[p 134]{esopcallneed}:
$$\begin{array}{l@{\qquad}l}
e\in \mathit{se}::=x\mid\lambda x.e\mid e\, e
&\hat{A}\in \mathit{s\hat{A}}::=\ContextHole\mid A[\hat{A}]\,e
\\
v\in \mathit{sv} ::=\lambda x.e
& \check{A}\in \mathit{s\check{A}}::=\ContextHole\mid A[\lambda x. \check{A}]\,e
\\
a\in \mathit{sa}::=A[v]
&E\in \mathit{sE}::=\ContextHole\mid E\, e\mid A[E]\mid \hat{A}[A[\lambda x.\check{A}[E[x]]]E]
\\
A\in sA::=\ContextHole\mid A[\lambda x.A]\, e&
\qquad\qquad\qquad\qquad\qquad\quad\mathrm{where}\hat{A}[\check{A}]\in A\\
\end{array} $$
Each constraint is added sequentially and the least set of objects satisfying them is recalculated.
Where the value of a set a metavariable can range over is recalculated and it is referenced in another rule, the set that rule applies to is recalculated with a new value.
For example, initially $a\in sa=\emptyset$, but when $A::=\ContextHole$ is read it triggers a recalculation of $A[v]$ so $ sa=sv$. Then when 
$A::=A[\lambda x.A]\, e$ is read, first it triggers a recalculation of $A$ so $sA=\{\ContextHole \}
\cup
\AllSuchThat{\ArrangementEquivClass{P_{\lambda A}\,P_e}}{\mtnormsf{ptr}(e)=P_e\wedge \mtnormsf{ptr}(\ArrangementEquivClass{\lambda x.\ContextHole })=P_{\lambda A}}$
 then it triggers a recalculation of $a$ so 
 $sa=v\,\cup\AllSuchThat{\ArrangementEquivClass{P_{\lambda A[v]}\,P_e}}{\mtnormsf{ptr}(e)=P_e\wedge \mtnormsf{ptr}(\ArrangementEquivClass{\lambda x.P_v })=P_{\lambda A[v]}\wedge \mtnormsf{ptr}(v)=P_v}$
then $a$ won't trigger recalculations, but we have a recalculation on $A$ waiting.
Let $f(\mathit{se},\mathit{sv},\mathit{sa},\mathit{sA},\mathit{s\hat{A}},\mathit{s\check{A}},\mathit{sE})$ take $\Tuple{\mathit{se},\mathit{sv},\mathit{sa},\mathit{sA},\mathit{s\hat{A}},\mathit{s\check{A}},\mathit{sE}}$ when a recalculation is triggered to their values after a recalculation is performed.
Let $<$ be an relation on 
$\Tuple{\mathit{se},\mathit{sv},\mathit{sa},\mathit{sA},\mathit{s\hat{A}},\mathit{s\check{A}},\mathit{sE}}$ such that 
$\Tuple{\mathit{se}^1,\mathit{sv}^1,\mathit{sa}^1,\mathit{sA}^1,\mathit{s\hat{A}}^1,\mathit{s\check{A}}^1,\mathit{sE}^1}
<\Tuple{\mathit{se}^2,\mathit{sv}^2,\mathit{sa}^2,\mathit{sA}^2,\mathit{s\hat{A}}^2,\mathit{s\check{A}}^2,\mathit{sE}^2}$ iff 
$\mathit{se}^1\subset \mathit{se}^2$ or $ \mathit{sv}^1\subset \mathit{sv}^2$ or $ \mathit{sa}^1\subset \mathit{sa}^2$ or 
$ \mathit{sA}^1\subset \mathit{sA}^2$ or $ \mathit{s\hat{A}}^1\subset \mathit{s\hat{A}}^2$ or $\mathit{s\check{A}}^1\subset \mathit{s\check{A}}^2$ or $ \mathit{sE}^1\subset \mathit{sE}^2$.
We observe that each set out of $\Tuple{\mathit{se},\mathit{sv},\mathit{sa},\mathit{sA},\mathit{s\hat{A}},\mathit{s\check{A}},\mathit{sE}}$ either gets new elements added to it or remains the same every time a recalculation is triggered and is bounded above by $\ObjectSet$. 
We can therefore take the least fixed point on $f$ satisfying all of these constraints.

If the side condition on $E$ were to re-trigger the calculation on $A$, $\hat{A}$ or $\check{A}$ we would have to be able to check that the new side condition produced by this recalculation could not effect any $E$ previously added.
This grammar relies on an assignation of values to $x$, otherwise all of its sets are either $\emptyset$ or $\ContextHole$.
For a given equivalence (e.g. $\equiv_\alpha$ or $\equiv_{\Arrangement}$) this grammar may define a different collection of $\Tuple{\mathit{se},\mathit{sv},\mathit{sa},\mathit{sA},\mathit{s\hat{A}},\mathit{s\check{A}},\mathit{sE}}$.

We add the reduction rule for this Grammar which is the least $\mathcal{R}$ satisfying:
$$
\begin{array}{l@{\quad}l}
\hat{A}[A_1[\lambda x.\check{A}[E[x]]]A_2[v]]\GroundRelation{\mathcal{R}} {}(\hat{A}[A_1[A_2[\MetaSubst{(\check{A}[E[x]])}{x}{v}]]])&\quad\mathrm{where\,}\hat{A}[\check{A}]\in A
\end{array}$$



\end{document}